\documentclass[twocolumn, trackchanges]{aastex62}

\usepackage{amsmath}
\usepackage{amssymb}
\usepackage{verbatim}
\usepackage{subfigure}
\usepackage{hyperref}
\usepackage{booktabs}
\usepackage{microtype}

\graphicspath{{figs/}}

\newcommand{\E}[1]{\ensuremath{\times 10^{#1}}}

\newcommand{\like}{\ensuremath{\mathcal{L}}}
\newcommand{\normal}{\ensuremath{\mathcal{N}}}
\newcommand{\uniform}{\ensuremath{\mathcal{U}}}
\newcommand{\sersic}{S\'{e}rsic}

\newcommand{\kms}{\ensuremath{\ \mathrm{km \ s}^{-1}}}
\newcommand{\eV}{\ensuremath{\ \mathrm{eV}}}
\newcommand{\vir}{\ensuremath{\mathrm{200c}}}
\newcommand{\sol}{\ensuremath{\mathrm{sol}}}
\newcommand{\abg}{\ensuremath{\alpha\beta\gamma}}

\defcitealias{vanDokkum2019a}{Paper I}
\defcitealias{Planck2018}{Planck Collaboration 2018}

\shorttitle{Fuzzy DM in Dragonfly~44}
\shortauthors{Wasserman et al.}

\begin{document}

\title{Spatially Resolved Stellar Kinematics of the Ultra-Diffuse Galaxy Dragonfly~44: \\
  II. Constraints on Fuzzy Dark Matter}

% alternative titles:
% What the Fuzz?
% Ultra-light DM in an Ultra-Diffuse Galaxy

\correspondingauthor{Asher Wasserman}
\email{adwasser@ucsc.edu}

\author[0000-0003-4235-3595]{Asher Wasserman}
\affil{Department of Astronomy \& Astrophysics, University of California Santa Cruz, Santa Cruz, CA 95064, USA}
\author{Pieter van Dokkum}
\affil{Astronomy Department, Yale University, New Haven, CT 06511, USA}
\author[0000-0003-2473-0369]{Aaron J. Romanowsky}
\affil{Department of Physics \& Astronomy, San Jos\'{e} State University, One Washington Square, San Jose, CA 95192, USA}
\affil{University of California Observatories, 1156 High Street, Santa Cruz, CA 95064, USA}
\author{Jean Brodie}
\affil{Department of Astronomy \& Astrophysics, University of California Santa Cruz, Santa Cruz, CA 95064, USA}
\affil{University of California Observatories, 1156 High Street, Santa Cruz, CA 95064, USA}
\author{Shany Danieli}
\affil{Astronomy Department, Yale University, New Haven, CT 06511, USA}
\author{Duncan A. Forbes}
\affil{Centre for Astrophysics and Supercomputing, Swinburne University,
Hawthorn, VIC 3122, Australia}
\author{Roberto Abraham}
\affil{Department of Astronomy \& Astrophysics, University of Toronto, 50 St. George Street, Toronto, ON M5S 3H4, Canada}
\author{Christopher Martin}
\affil{Cahill Center for Astrophysics, California Institute of Technology, 1216 East California Boulevard, Mail code 278-17, Pasadena, CA 91125, USA}
\author{Matt Matuszewski} 
\affil{Cahill Center for Astrophysics, California Institute of Technology, 1216 East California Boulevard, Mail code 278-17, Pasadena, CA 91125, USA}
\author{Alexa Villaume}
\affil{Department of Astronomy \& Astrophysics, University of California Santa Cruz, Santa Cruz, CA 95064, USA}
\author{John Tamanas}
\affil{Department of Physics, 1156 High St., University of California Santa Cruz, Santa Cruz, CA 95064, USA}
\affil{Santa Cruz Institute for Particle Physics, 1156 High St., Santa Cruz, CA 95064, USA}
\author{Stefano Profumo}
\affil{Department of Physics, 1156 High St., University of California Santa Cruz, Santa Cruz, CA 95064, USA}
\affil{Santa Cruz Institute for Particle Physics, 1156 High St., Santa Cruz, CA 95064, USA}

\begin{abstract}
  Given the absence of directly detected dark matter (DM) as weakly interacting massive particles, there is strong interest in the possibility that DM is an ultra-light scalar field, here denoted as ``fuzzy'' DM.
  Ultra-diffuse galaxies, with the sizes of giant galaxies and the luminosities of dwarf galaxies, have a wide range of DM halo masses, thus providing new opportunities for exploring the connections between galaxies and their DM halos.
  Following up on new integral field unit spectroscopic observations and dynamics modeling of the DM-dominated ultra-diffuse galaxy Dragonfly~44 in the outskirts of the Coma Cluster, we present models of fuzzy DM constrained by the stellar dynamics of this galaxy.
  We infer a scalar field mass of $\sim 3\E{-22} \eV$, consistent with other constraints from galaxy dynamics but in tension with constraints from Ly$\alpha$ forest power spectrum modeling.
  While we are unable to statistically distinguish between fuzzy DM and ``normal'' cold DM models, we find that the inferred properties of the fuzzy DM halo satisfy a number of predictions for halos in a fuzzy DM cosmology.
  In particular, we find good agreement with the predicted core size--halo mass relation and the predicted transition radius between the quantum pressure-dominated inner region and the outer halo region.
\end{abstract}

\keywords{galaxies: halos --- galaxies: individual (Dragonfly~44) --- galaxies: kinematics and dynamics --- cosmology: dark matter}

\section{Introduction}\label{sec:introduction}

The concordant cosmological model of dark energy plus cold dark matter ($\Lambda$CDM) has had remarkable successes in describing the large scale structure of the universe (e.g., \citealt{Tegmark2006}; \citetalias{Planck2018}).
However, there have been a number of small scale challenges to this picture concerning the inner density structure of dark matter (DM) halos and the relative numbers of subhalos \citep[e.g.,][and references therein]{Weinberg2015, Bullock2017}.
Many authors have proposed solutions to these problems that involve a more detailed treatment of the baryonic physics of galaxy formation \citep[e.g.,][]{Pontzen2012, Martizzi2013, Schaller2015}.
Furthermore, given the continued absence of directly detected DM particles \citep{MarrodanUndagoitia2016, Akerib2017, Aprile2018}, attempts to explain these astrophysical discrepancies with modifications of the physics of DM have become increasingly appealing.
Frequently considered modifications include allowing for self-interactions \citep[SIDM; e.g.,][]{Carlson1992, Rocha2013, Wittman2018} and increasing the DM temperature at the time of thermal decoupling \citep[Warm DM; e.g.,][]{Davis1981, Lovell2017, Bozek2019}.
For overviews of the intersection of astrophysics and particle physics searches for DM, we refer readers to the reviews of \cite{Bertone2005}, \cite{Profumo2017}, and \cite{Buckley2018}.

One promising class of models posits that the DM particle is an extremely low-mass ($\ll 1 \eV$)\footnote{For particle masses, we use the convention that $c = 1$, giving mass and energy the same physical dimensions.} spin-0 boson  (i.e., a scalar field) manifesting quantum mechanical wave-like behavior on astrophysical scales \citep[$\sim$kpc;][]{Colpi1986, Lee1996, Hu2000, Matos2009, Hui2017}.
Axions, a proposed solution to the strong Charge-Parity (CP) problem in particle physics \citep{Peccei1977, Dine1981}, are a well-motivated class of models that provide one such candidate DM particle.
There are a variety of names for these DM models: ultra-light axion DM, scalar field DM, Bose-Einstein condensate DM, wave DM, or fuzzy DM.
Here we adopt the term Fuzzy Dark Matter (FDM) for ultra-light ($m \sim 10^{-22} \eV$) non-thermal (i.e., restricted to the ground state) models lacking self-interaction.
We refer to the mass of the DM scalar field in this model in its dimensionless form as $m_{22} = m / 10^{-22} \eV$.

We note that for any model in which an ultra-light scalar field is the dominant contributor to DM, its production mechanism must necessarily be non-thermal \citep{Marsh2016}, in contrast with the thermal production of weakly interacting massive particles in the standard CDM cosmology \citep{Bringmann2007}.
Thermal production of such a low mass of DM would lead to hot (i.e. ultra-relativistic) DM, in conflict with observations of the matter power spectrum and the cosmic microwave background (CMB; e.g., \citealt{Viel2005}).
For a broad overview of FDM cosmologies, we refer interested readers to \cite{Marsh2016} and \cite{Hui2017}.

The salient phenomena associated with FDM cosmologies are a cutoff in the halo mass function below $\sim 10^{9}~M_\odot$, and distinct density cores in the inner $\sim 1$~kpc of DM halos, with a lighter scalar field mass resulting in a higher halo mass cutoff and a more massive inner core \citep{Hu2000}.
This cutoff in the halo mass distribution implies less correlation of structure on smaller scales and the delayed formation of galaxies relative to CDM.
The measured CMB and galaxy power spectra imply that, if FDM makes up the majority of DM in the universe, $m_{22}$ must be $\gtrsim 10^{-3}$ \citep{Hlozek2015}.
Constraints from the Ly$\alpha$ forest power spectrum imply that $m_{22} \gtrsim 1$, with some models excluding scalar field masses up to $m_{22} \sim 30$ \citep{Armengaud2017, Nori2019}.
Complementary constraints on FDM models from both high redshift galaxy luminosity functions and the Milky Way satellite luminosity function are also consistent with $m_{22} \gtrsim 1$ \citep{Bozek2015, Schive2016, Nadler2019}.

% \note{21 cm constraints? pulsar timing?}

The stellar dynamics of nearby galaxies offer further opportunities to test FDM models.  
The inner density structures of DM halos that form in an FDM cosmology follow a stationary wave, or soliton, solution to the Schr\"{o}dinger--Poisson equation \citep{Schive2014, Marsh2015}.
In the outer region the halo density profile transitions to a normal CDM halo profile \citep[e.g., a Navarro-Frenk-White (NFW) profile,][]{Navarro1997}.
The sizes of these cores are predicted to scale inversely with halo mass, while the symmetry of the soliton solution requires the core density to scale inversely with the core size \citep{Schive2014}.
Higher mass halos are therefore predicted to have smaller but denser cores.

Many previous studies of FDM density profiles in galaxies have focused on either dwarf spheroidal (dSph) or ultra-faint dwarf (UFD) galaxies \citep[e.g.,][]{Lora2014, Marsh2015, Chen2017, Gonzalez-Morales2017}, as their high dynamical mass-to-light ratios minimize the impact of systematic assumptions about the stellar mass distribution.
Studies have generally found $m_{22} \sim 1$ (within a factor of a few), in slight tension with the Ly$\alpha$ constraints.
\cite{Calabrese2016} found that the stellar kinematics of two UFDs were consistent with $m_{22} \sim 4$, though they noted the lack of kinematic measurements outside of the inferred core radius.
More recently, \cite{Marsh2018} applied the stochastic density fluctuation model of \cite{El-Zant2016} to study how FDM would cause dynamical heating of the star cluster in the UFD Eridanus II.
They argued that the survival of the EriII star cluster implies $m_{22} \gtrsim 1000$, whereas the existence of EriII itself implies $m_{22} \gtrsim 10$.

Looking toward more massive galaxies to probe FDM scaling relations presents increasing difficulties in disentangling the baryonic and dark mass components.
In the halo mass range of $10^{10}$ - $10^{11} \ M_\odot$, low surface brightness (LSB) galaxies have proven to be the most amenable to analysis.
\cite{Bernal2018} modeled the rotation curves of 18 LSBs, and their results favored a lower value of $m_{22} \sim 0.05$ (though see \citealt{Bar2019} for a discussion of the impact of the baryons on the FDM density structure).

% \note{MW constraints? \url{http://adsabs.harvard.edu/abs/2018arXiv180708153D}}
% also Nadler+2019

With the discovery of a vast population of even lower surface brightness ``ultra-diffuse'' galaxies \citep[UDGs;][]{Dokkum2015,Mihos2015,Koda2015}, we now have more opportunities to test FDM in a broader range of galaxy masses and environments.
The Coma Cluster UDG Dragonfly~44 was shown to have a large stellar velocity dispersion, corresponding to a DM halo with a mass on the order of that of the Milky Way \citep{Dokkum2016}.
In a companion paper, \cite{vanDokkum2019a} \citepalias[hereafter][]{vanDokkum2019a}, we present new spatially-resolved spectroscopy of Dragonfly~44, confirming that the potential of the galaxy is indeed dominated by DM.  
In this work, we address the question of whether or not the dynamics of Dragonfly~44 are consistent with FDM.

Throughout this work we assume the \citetalias{Planck2018} values of relevant cosmological parameters, including $H_0 = 67.66$~km~s$^{-1}$~Mpc$^{-1}$ and $\Omega_m = 0.3111$.

In Section~\ref{sec:data} we summarize the photometric and spectroscopic data for Dragonfly~44. 
We describe the Jeans modeling formalism and mass modeling assumptions in Section~\ref{sec:methods}.  
In Section~\ref{sec:results} we present our derived constraints on FDM models, and we place our results in context with other FDM studies in Section~\ref{sec:discussion}.

\section{Data}\label{sec:data}

Readers interested in a detailed description of the spectroscopic observations, data reduction, and kinematic extraction are referred to \citetalias{vanDokkum2019a}; here we provide a brief summary of the observational data for Dragonfly~44.
We adopt a standard distance to Coma of 100 Mpc for the galaxy, which has an associated distance modulus $m -M = 35$ and an angular distance conversion factor of $0.485$~kpc~arcsec$^{-1}$.

Using the $V_{606}$ \emph{HST} WFC3/UVIS imaging data presented by \cite{vanDokkum2017a},
we modeled the stellar light of Dragonfly~44 with a \sersic{} surface brightness profile, deriving a total luminosity of $L_{V} = 2.33\E{8} \ L_{\odot, V}$, a major-axis effective radius of $R_e = 4.7$ kpc, a \sersic{} index of $n = 0.94$, and an axis ratio of $b/a = 0.68$.
For our modeling purposes, we adopt the circularized effective radius of $R_{e,\mathrm{circ}} = R_e \sqrt{b/a} = 3.87$ kpc.

We obtained integral field unit (IFU) spectroscopy of Dragonfly~44 with the Keck Cosmic Web Imager (KCWI) in the first half of 2018, with 17 hours of exposure time on target and an additional 8 hours on sky.
We used the medium slicer with the BM grating, yielding a field-of-view of $16\arcsec \times 20\arcsec$ and a spectral resolution of $R \sim 4000$.

For reducing the data to rectified, wavelength calibrated cubes, we used the public Keck-maintained pipeline, KDERP\footnote{\url{https://github.com/Keck-DataReductionPipelines/KcwiDRP}}.
We aligned the individual science exposures by fitting a 2D model of the flux from the \emph{HST} imaging data and interpolating to a common spatial grid with a spatial resolution of $\sim 1.2\arcsec$.
We subtracted the sky spectrum using a principle component analysis technique -- see \citetalias{vanDokkum2019a} for further details.
The final signal-to-noise ratio in the optimally-combined spectrum was 48 per pixel or 96 \AA$^{-1}$.

We extracted spectra in nine elliptical apertures following the isophotes of the galaxy.
We modeled the stellar kinematic line-of-sight velocity distribution (LOSVD) as a fourth-order Gauss--Hermite function, and we fitted the LOSVD in each of these apertures by convolving it with both a high-resolution template spectrum of a synthetic stellar population and the instrumental line profile (including a wavelength-dependent resolution).
From varying the ages and metallicities of the chosen stellar population template, we found the most likely values for an age of 10~Gyr and a metallicity of $[\mathrm{Fe/H}] = -1.25$.
For each spectrum we found the best fitting central velocity and higher-order (second, third, and fourth) moments of the LOSVD using a Markov Chain Monte Carlo (MCMC) simulation.

The radius of a given aperture is defined as the flux-weighted average pixel radius.
There is little evidence for rotational motion in Dragonfly~44, with $v/\sigma \lesssim 0.25$ along the minor axis and $v/\sigma \lesssim 0.1$ along the major axis.  We computed the effective rms velocity within each aperture as $v_\mathrm{rms}^2 = (v - v_\mathrm{sys})^2 + \sigma^2$.

\section{Dynamical Modeling}\label{sec:methods}

We use the spherical Jeans modeling formalism presented in \cite{Wasserman2018}, using an updated, publicly available modeling code\footnote{\url{http://github.com/adwasser/Slomo.jl}}.
Under the assumptions of dynamical equilibrium and spherical symmetry, the model predicts the LOS velocity dispersion as a function of projected galactocentric radius.
See \cite{Hayashi2019} for a discussion of the systematic uncertainty associated with applying spherical models to non-spherical systems.
The main components of the model are the mass profile, $M(r)$, the tracer volume density profile, $\nu(r)$, and the orbital anisotropy profile of the tracers, $\beta_\mathrm{ani}(r) = 1 - \sigma^2_t / \sigma^2_r$, where $\sigma_t$ and $\sigma_r$ are the tangential and radial components of the velocity dispersion.  
We can compute the mean squared LOS velocity as
\begin{equation}
  \label{eq:jeans}
  \sigma_\mathrm{los}^2(R) = \frac{2G}{I(R)} \int_R^\infty K_\beta\left(r, R\right) \nu(r) M(r) \frac{dr}{r}
\end{equation}
where $I(R)$ is the tracer surface density profile and $K_\beta(r, R)$ is the anisotropy projection kernel.
For our adopted constant anisotropy profile, the functional form of this projection kernel is given by \cite{Mamon2005}, equation A16.

We set the stellar tracer density distribution to follow the \sersic{} distribution of the star light.
We assume that the stellar mass distribution follows the same \sersic{} luminosity distribution used for the tracers, with the local stellar mass density given by the spatially-invariant stellar mass-to-light ratio, $\Upsilon_*$, multiplied by the stellar luminosity density.

\subsection{Halo Models}\label{sec:halo}

For the DM halo, we construct a flexible double power law model with a soliton core.
A generalized form of the Navarro--Frenk--White (NFW) model \citep{Navarro1997} is given by
\begin{equation}
  \rho_{\abg}(r) = \rho_s \left(\frac{r}{r_s}\right)^{-\gamma} \left(1 + \left(\frac{r}{r_s}\right)^{\alpha}\right)^{(\gamma - \beta)/\alpha}
\end{equation}
where $\rho_s$ is the scale density, $r_s$ is the scale radius, $\gamma$ is the negative inner log slope, $\beta$ is the negative outer log slope, and $\alpha$ controls the sharpness of the transition between the two slopes \citep{Hernquist1990, DiCintio2014a}.
For $(\alpha, \beta, \gamma) = (1, 3, 1)$, this is the typical NFW profile, which we assume to be an appropriate approximation for CDM halos in the absence of baryonic effects or FDM cores.

The inner soliton core region from FDM has the density profile
\begin{equation}
  \rho_\mathrm{soliton}(r) = \rho_\sol{} \left(1 + \left(\frac{r}{r_\sol{}}\right)^2\right)^{-8}
\end{equation}
where $\rho_\sol{}$ and $r_\sol{}$ are the soliton scale density and scale radius, respectively \citep{Marsh2015, Robles2019}.
Note that we use a slightly different definition of the soliton radius than \cite{Schive2014} and \cite{Robles2019}; their core radius, $r_c$, refers to the radius where the density has fallen to half of the central density, and it is equivalent to $0.3017 \ r_\sol{}$.
In addition to eliminating a numeric constant from the equations, our choice of definition for the soliton radius makes the ratio of the transition radius to the soliton radius near unity (see Section~\ref{sec:discussion-transition}).

From the symmetry of the soliton solution, the soliton scale density and radius are related to the scalar field mass as
\begin{equation}\label{eq:solscaling}
  \frac{\rho_\sol{}}{M_\odot \mathrm{kpc}^{-3}} = 8.755\E{6} \ h^{-2} \ m_{22}^{-2} \left(\frac{r_\sol{}}{\mathrm{kpc}}\right)^{-4}
\end{equation}
where $h$ is the Hubble parameter in units of $100$ Mpc~\kms, and $m_{22}$ is the scalar field mass in units of $10^{-22} \eV$ \citep{Marsh2015}.

We match the inner soliton profile with the outer $\abg$ profile at the transition radius, $r_t$, by finding the root of the function corresponding to the difference between the two profiles.
This guarantees that the density profile,
\begin{equation}\label{eq:density}
  \rho(r) =
  \begin{cases}
    \rho_\mathrm{soliton}(r) & r < r_t \\
    \rho_{\alpha\beta\gamma}(r) & r \geq r_t \ ,\\
  \end{cases}
\end{equation}
is a continuous function, and the transition radius is thus fixed for a given set of outer halo and soliton parameters.
We reject any model that fails to converge due to the inner profile being less dense than the outer profile at all radii.
The transition radius is found in simulations to be a factor of a few times the core radius of the soliton, and the transition between the soliton and normal CDM profiles is sharp \citep{Schive2014, Mocz2017}.
While FDM halo density profiles are continuous, their density derivatives are not.

The enclosed mass in the $\abg$ model is 
\begin{equation}
  \label{eq:abg_mass}
  M_{\abg}(r) = \frac{4\pi\rho_s r_s^3}{\omega} \left(\frac{r}{r_s}\right)^{\omega}  \ _2F_1\left[\frac{\omega}{\alpha}, \frac{\beta - \gamma}{\alpha}, 1 + \frac{\omega}{\alpha}; -x^\alpha \right]
\end{equation}
where $\omega = 3 - \gamma$ and $_2F_1$ is the hypergeometric function.  
For the limiting case of the NFW profile, this simplifies to
\begin{equation}
  \label{eq:nfw_mass}
  M_\mathrm{NFW}(r) = 4\pi\rho_s r_s^3 \left[\ln\left(1 + \frac{r}{r_s}\right) - \frac{r}{r_s + r}\right] \ .
\end{equation}

The enclosed mass of the soliton has an analytic form\footnote{The existence of such an analytic form was noted by \cite{Marsh2015}, but the derivation of this profile was left as an exercise to the reader.} and is given by
\begin{align}
  M_\mathrm{soliton}(r) 
  &= \int_0^r 4\pi \tilde{r}^2 \rho_\mathrm{soliton}(\tilde{r}) d\tilde{r} \\
  &= 4\pi \rho_\sol{} r_\sol{}^3 \int_0^{r / r_\sol{}} x^2 \left(1 + x^2\right)^{-8} \ dx \nonumber \\
  &= 4\pi \rho_\sol{} r_\sol{}^3 \int_0^\theta \tan^2(\theta) \sec^{-16}(\theta) \sec^2(\theta) \ d\theta \nonumber \\
  &= 4\pi \rho_\sol{} r_\sol{}^3 \int_0^\theta \sin^2(\theta) \cos^{12}(\theta) \ d\theta \nonumber
\end{align}
where in the second-to-last line we have used the trigonometric substitution $r / r_\sol{} = \tan(\theta)$.
The integral in the last line can then be iteratively integrated by parts, yielding the following solution.

\begin{align}\label{eq:solmass_integral}
  M_\mathrm{soliton}(r) = M_\sol \frac{1}{K} \Big[k_0\theta + \sum_{i = 1}^7 k_i\sin(2i\theta) \Big]
\end{align}
where $M_\sol = 4\pi\rho_\sol{} r_\sol{}^3$, $K = 1720320$ and the other constant factors are given in the table below.

\begin{table}[h!]
  \centering
  \begin{tabular}{llllllll}
    \toprule
    $k_0$ & $k_1$ & $k_2$ & $k_3$ & $k_4$ & $k_5$ & $k_6$ & $k_7$ \\ 
    \midrule
    27720 & 17325 & $-$1155 & $-$4235 & $-$2625 & $-$903 & $-$175 & $-$15 \\ 
    \bottomrule
  \end{tabular}
  \caption{Coefficients for the analytic solution to the soliton enclosed mass profile (Equation~\ref{eq:solmass_integral}).}
\end{table}

From Equation~\ref{eq:solscaling}, we can also express $M_\sol$ as
\begin{equation}
  M_\sol = 1.1\E{8} \ M_\odot \ h^{-2} \ m_{22}^{-2} \ \left(\frac{r_\sol}{\mathrm{kpc}}\right)^{-1} \ .
\end{equation}

Our generic halo mass profile is then given by
\begin{equation}
  M(r) = 
  \begin{cases}
    M_\mathrm{soliton}(r) & r < r_t \\
    \Delta M_{\abg}(r) + M_\mathrm{soliton}(r_t) & r \geq r_t \\
  \end{cases}
\end{equation}
where $\Delta M_{\abg}(r) = M_{\abg}(r) - M_{\abg}(r_t)$.

We parameterize the halo with the virial mass and concentration, using the ``200c'' convention such that the virial radius is given by the relation
\begin{align}
  M(r_\vir) = 200\rho_\mathrm{crit} \frac{4\pi}{3} r_\vir^3
\end{align}
and $c_\vir = r_\vir / r_{-2}$.  
Note that here we use the convention that the halo concentration is given by the radius where the halo log slope is equal to $-2$.
This is related to the halo scale radius as
\begin{equation}
  r_{-2} = \left(\frac{2 - \gamma}{\beta - 2}\right)^{1/\alpha} r_s \ .
\end{equation}

Generally speaking, we must be careful in our definition of the halo virial mass and concentration.  
Since the soliton core contributes to the mass of a halo, the outer halo density and radius scale parameters for a FDM halo of a given virial mass and concentration are necessarily different than those for a normal CDM halo.

However from the predicted scaling relation between soliton core mass and halo mass, we would expect a $10^{10} \ M_\odot$ halo to have $\lesssim 1\%$ of its mass locked up in the soliton core, with this fraction decreasing with increasing halo mass \citep{Robles2019}.
Thus given the expected halo mass range of Dragonfly~44 of $\sim 10^{11}-10^{12} \ M_\odot$, we assume that the differences in the outer halo scale parameters in the FDM and CDM models at fixed halo mass and concentration are negligible.
We later verify the validity of this assumption by comparing the inferred virial mass with one computed from the posterior mass profile, finding a negligible difference.

This generic double-power law plus soliton halo model has eight free parameters ($\beta_\mathrm{ani}$, $M_\vir$, $c_\vir$, $\alpha$, $\beta$, $\gamma$, $m_{22}$, $r_\sol$) and it would be poorly constrained by the available kinematic data.  
Thus, we consider the following constraints.

We impose a prior on $c_\vir$ by using the halo mass--concentration relation (HMCR) from \cite{Diemer2015}.  
Practically this is accomplished by sampling both $M_\vir$ and $c_\vir$, then using a log-normal prior on $c_\vir$ whose mean is the HMCR prediction conditioned on the sampled $M_\vir$, and whose scatter is 0.16 dex.

We consider two possibilities for the $\abg$ slope parameters.  
First, in the limit of no baryonic effects, we assume the outer halo follows an NFW profile with $(\alpha, \beta, \gamma) = (1, 3, 1)$.  
Alternatively assuming that baryonic feedback -- such as cycles of bursty star formation -- plays an important role,  we use the halo scaling relations from the hydrodynamics simulations of \cite{DiCintio2014a}, which map variation in $\log(M_* / M_\mathrm{vir})$ to $\alpha$, $\beta$, and $\gamma$ (see their Equation~3).
For Dragonfly~44, this results in a shallower CDM halo, with $\gamma \sim 0.3$.

% In \citetalias{vanDokkum2019a} we presented the results of modeling the Dragonfly~44 stellar dynamics using two CDM halo models, with the NFW and $\abg$ profiles.
To summarize, in addition to the CDM halo models described in \citetalias{vanDokkum2019a}, we have added two halo models by including the soliton core component from the FDM model, with both NFW and $\abg$ outer halo profiles.

\begin{figure*}
  \centering
  \includegraphics[width=0.8\linewidth]{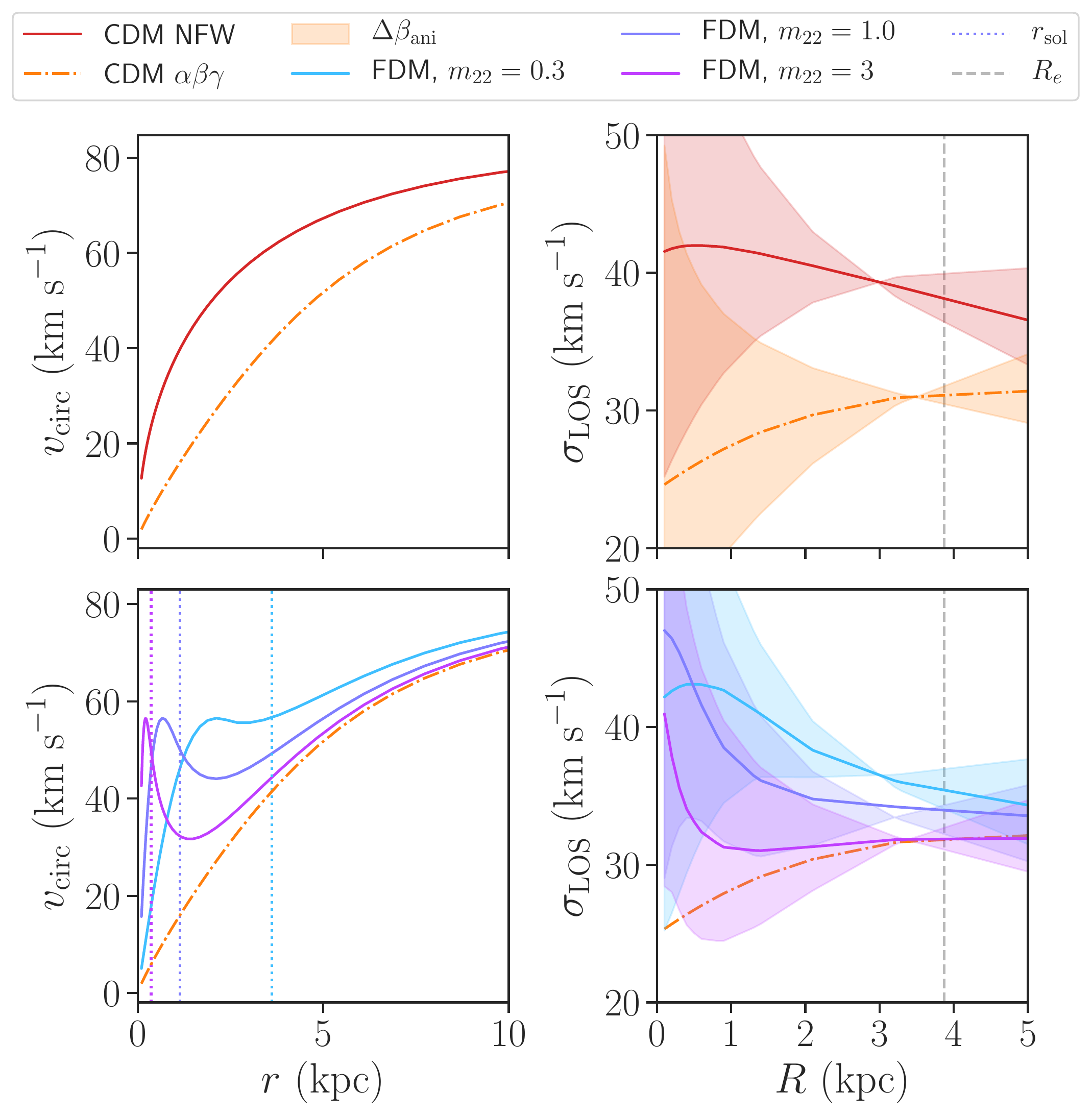}
  \caption{
    Illustration of mass models and their associated velocity dispersion profiles for different halo models described in Section~\ref{sec:halo}.
    The top panels show CDM models with $\log_{10} M_\vir / M_\odot = 11$, $c_\vir = 10.5$, and $r_s = 9.3$ kpc.
    The red solid line shows a cuspy NFW halo and the orange dot-dashed line shows a cored $\alpha\beta\gamma$ halo.
    The bottom panels show FDM halos with an outer $\alpha\beta\gamma$ halo profile (plotted again for comparison) for a range of possible values of $m_{22}$.
    The left-hand panels show the circular velocity profile associated to the halo, while the right-hand panels show the line-of-sight velocity dispersion profile.
    The range of orbital anisotropy values (from $\beta_\mathrm{ani} = -1$ to $0.5$) is shown by the shaded region, with the line indicating the isotropic ($\beta_\mathrm{ani} = 0$) profile.
    Tangentially-biased profiles ($\beta_\mathrm{ani} < 0$) generally display velocity dispersion profiles that increase with radius, while radially-biased profiles generally fall with radius.
    In the bottom left panel, the dotted lines show the expected soliton scale radius associated to each FDM halo (see Section~\ref{sec:discussion-core}).
    As the FDM scalar field mass gets larger, the profile approaches its CDM analogue, with the deviations occurring on increasingly smaller scales.
    FDM is more ``detectable'' for lower $m_{22}$ values where there is more mass in the soliton core.
    However, the projection of this mass profile into an observable velocity dispersion tends to wash out this signal (demonstrating the mass--anisotropy degeneracy).
    Furthermore even with a known anisotropy parameter, the FDM signal is degenerate with the inner DM slope (i.e., cored or cuspy).
  }\label{fig:cartoon}
\end{figure*}

Despite the constraints of the above assumptions, the task of inferring the properties of an FDM halo in Dragonfly~44 are substantial.
Figure~\ref{fig:cartoon} illustrates the difficulty by comparing velocity dispersion profiles from expected FDM halo models with their CDM counterparts.

\subsection{Bayesian Inference}

We use a Gaussian likelihood to model the stellar velocity dispersion data, $\sigma_i \pm \delta\sigma_i$ in apertures with projected galactocentric radii, $R_i$.
For a given halo model and model parameters, the predicted velocity dispersion, $\sigma_\mathrm{J}$, is modeled by Equation.~\ref{eq:jeans}.  
The log likelihood is thus

\begin{equation}
  \ln\like = \sum_i -\frac{1}{2} \left(\ln(2\pi\delta\sigma_i^2) + \left(\frac{\sigma_i - \sigma_\mathrm{J}(R_i)}{\delta\sigma_i}\right)^2 \right) \ .
\end{equation}

We use uniform priors over the log of the halo mass, scalar field mass, and soliton scale radius.  
For the orbital anisotropy, we use a uniform prior over the symmetrized anisotropy parameter, $\tilde{\beta}_\mathrm{ani} = -\log_{10}(1 - \beta_\mathrm{ani})$.
This ensures that radial and tangential orbits are given equal weight.
We use the HMCR as a prior for the concentration, as described in the previous section.
For the stellar mass-to-light ratio, we use a log-normal distribution with mean $\log_{10} \Upsilon_{*,V} = \log_{10}(1.5)$ and a scatter of $0.1$ dex.
Here the mean value chosen is typical of an old, low metallicity stellar population, while the chosen scatter matches that found by \cite{Taylor2011} from the GAMA survey.
We show a summary of these model parameters and our priors in Table~\ref{tab:parameters}.

For each halo model, we sample from our posterior probability distribution,
\begin{equation}
  \label{eq:bayes}
  \mathrm{Post}(\theta | (\sigma, \delta\sigma, R), \mathrm{Model}) \propto \frac{\like(\sigma | R, \mathrm{Model}, \theta)}{\mathrm{Prior}(\theta)}
\end{equation}
by using the affine-invariant ensemble MCMC algorithm of \cite{Goodman2010}.
We run chains of 128 walkers for 4000 iterations, rejecting the first 2000 iterations where the MCMC might not have converged.
We visually inspect the trace plots to verify that this is an adequate number of burn-in iterations.

\begin{table*}
  \centering
  \begin{tabular}{lllcccc}
    \toprule
    Parameter              & Unit                     & Prior                   & CDM + NFW                & FDM + NFW                & CDM + $\abg$             & FDM + $\abg$ \\ 
    \midrule
    $\log_{10} M_\vir$     & $M_\odot$                & $\uniform{}(7, 15)$     & $10.62^{+0.42}_{-0.30}$ & $10.64^{+0.41}_{-0.32}$ & $11.20^{+0.63}_{-0.63}$ & $11.16^{+0.58}_{-0.58}$ \\
    $\log_{10} c_\vir$     & --                       & HMCR                    & $1.00^{+0.19}_{-0.20}$  & $0.98^{+0.18}_{-0.19}$  & $0.98^{+0.13}_{-0.16}$  & $0.99^{+0.12}_{-0.14}$  \\
    $\log_{10} \Upsilon_*$ & $M_\odot / L_{\odot, V}$ & $\normal{}(0.176, 0.1)$ & $0.18^{+0.10}_{-0.10}$  & $0.18^{+0.10}_{-0.10}$  & $0.19^{+0.10}_{-0.10}$  & $0.18^{+0.10}_{-0.10}$  \\
    $\tilde{\beta}_\mathrm{ani}$        & --          & $\uniform{}(-1.5, 1.5)$ & $-0.24^{+0.10}_{-0.12}$  & $-0.44^{+0.22}_{-0.29}$  & $-0.05^{+0.08}_{-0.11}$  & $-0.16^{+0.15}_{-0.39}$  \\
    $\log_{10} m_{22}$     & $10^{-22} \eV$            & $\uniform{}(-3, 3)$     & --                       & $0.34^{+0.76}_{-0.25}$  & --                       & $0.51^{+0.62}_{-0.44}$  \\
    $\log_{10} r_\sol$     & kpc                      & $\uniform{}(-2, 1)$     & --                       & $-0.22^{+0.25}_{-0.34}$  & --                       & $-0.16^{+0.25}_{-0.26}$  \\
    \bottomrule
  \end{tabular}
  \caption{Model parameters for the two CDM halo models from \citetalias{vanDokkum2019a} and the two FDM halo models presented in this work. 
    The parameters are, from top to bottom, the halo virial mass, the halo concentration, the stellar mass-to-light ratio, the symmetrized anisotropy parameter ($\tilde{\beta}_\mathrm{ani} = -\log_{10}(1 - \beta_\mathrm{ani})$), the scalar field mass, and the soliton core radius.  
    Columns show the chosen parameterization, relevant units, the prior distribution, and posterior summaries for the four halo models.
    For the priors, $\mathcal{U}(\ell, u)$ denotes a uniform prior with lower bound $\ell$ and upper bound $u$, $\mathcal{N}(\mu, \sigma)$ denotes a Gaussian prior with mean $\mu$ and standard deviation $\sigma$, and HMCR refers to the halo mass--concentration relation prior (see Section~\ref{sec:halo}).
    Posterior distributions are summarized as the median of the distribution and the distance to the 16th and 84th percentiles.}
  \label{tab:parameters}
\end{table*}

\section{Results}\label{sec:results}

Table~\ref{tab:parameters} summarizes the posterior distributions for the different halo mass models.  
The full posterior distributions are shown as marginalized 1D and 2D histograms in Appendix~\ref{sec:appendix-posteriors}.

\begin{figure}
  \centering
  \includegraphics[width=\linewidth]{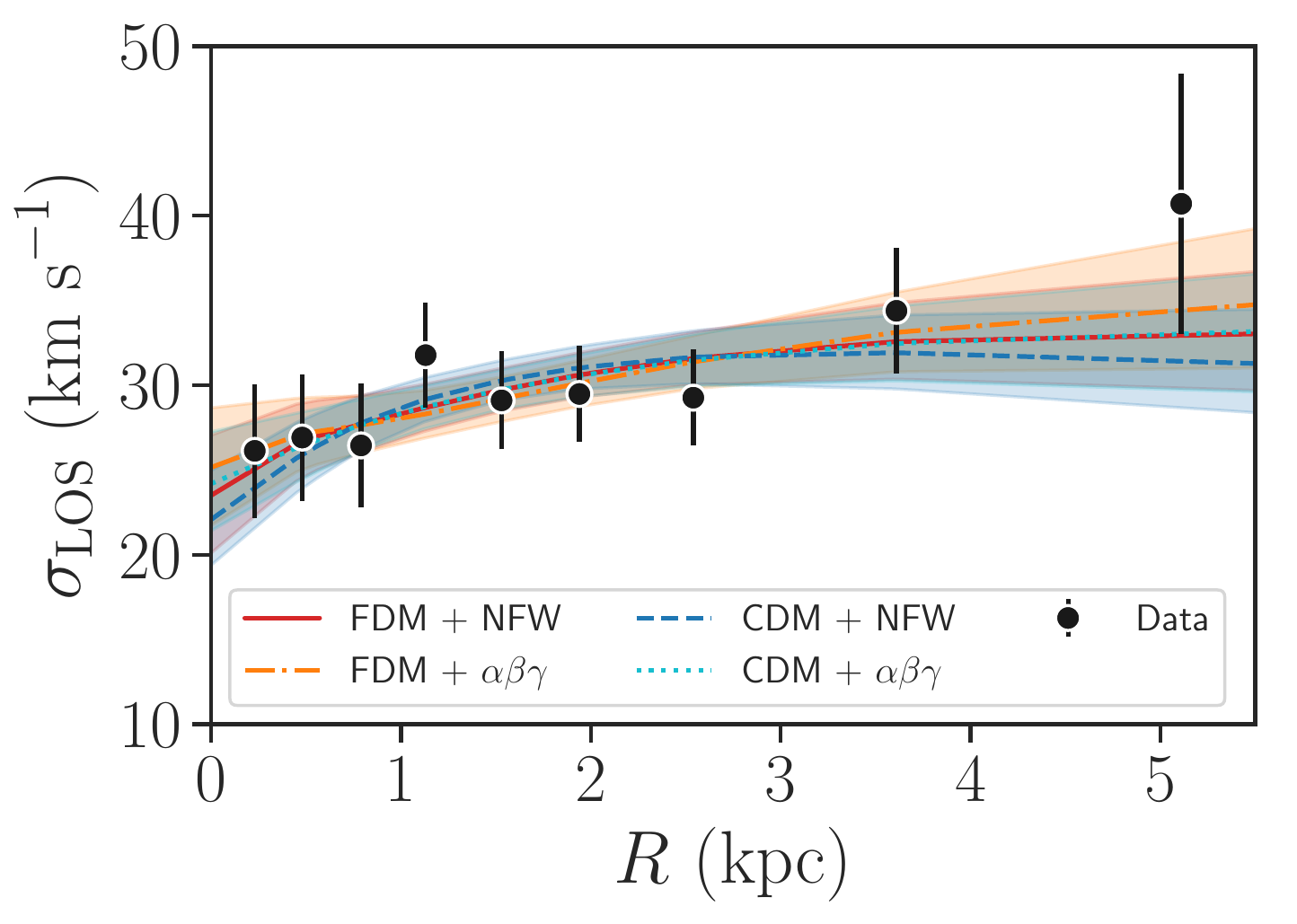}
  \caption{Posterior predictive checks on the velocity dispersion profiles for the FDM halo models compared with the kinematic observations, with the CDM halo models from \citetalias{vanDokkum2019a} shown for comparison.
    The red solid and orange dot-dashed lines show the FDM halo models for the NFW and $\abg$ outer profiles.
    The dark blue dashed and cyan dotted lines show the CDM halo models for the NFW and $\abg$ profiles.
    The shaded regions cover the 16th through 84th percentiles of the distribution.
    We see that all four models do an adequate job of recovering the general trend of the kinematic data.}
  \label{fig:sigma_ppc}
\end{figure}

We find that all models we consider are able to reproduce the observed velocity dispersion profile, as shown in Figure~\ref{fig:sigma_ppc}.  
We assess the relative quality of these models using leave-one-out cross validation \citep[LOO-CV;][]{Vehtari2015b, Piironen2017}, finding no significant differences in the goodness-of-fit of FDM models relative to the CDM models.
Translating the differences between models in their calculated LOO-CV information criteria into probabilities, we find that no model is more than $\sim 0.3$ times as likely as any other model to best describe the data.
In other words, the increase in goodness-of-fit from the FDM models is not enough to compensate for the increased model freedom (i.e., the additional model parameters).

\begin{figure}
  \centering
  \includegraphics[width=\linewidth]{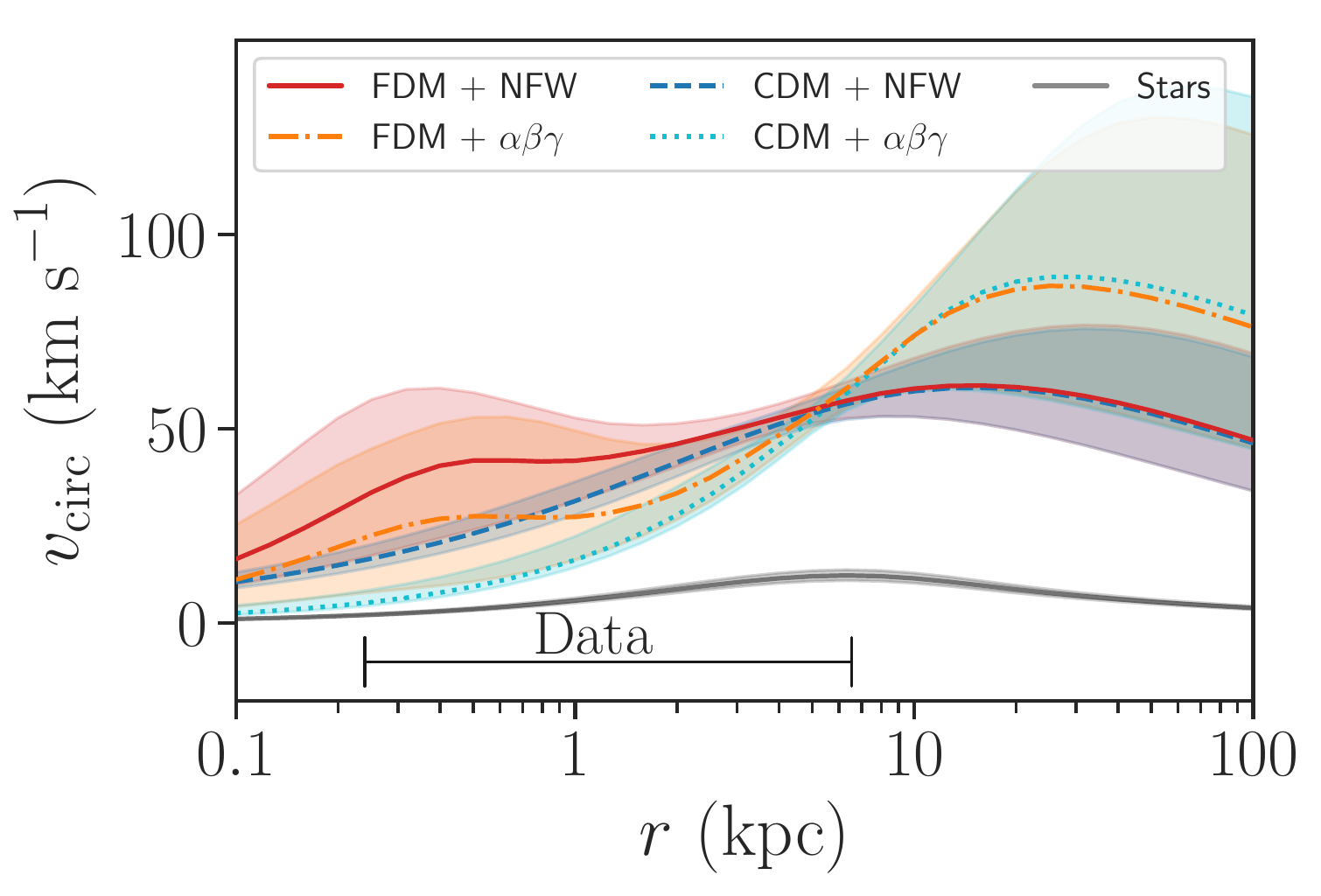}
  \caption{Circular velocity corresponding to the dynamical mass (DM + stars) for the FDM halo models, compared with their CDM halo counterparts.
    Note that these are profiles in de-projected (3D) radius, in contrast to the projected (2D) radial profiles shown in Figure~\ref{fig:sigma_ppc}.
    The bottom gray solid line shows the circular velocity profile corresponding to just the stellar mass for the NFW model.
    The black bar at the bottom indicates the spatial extent of the kinematic data.
    The dynamical mass within $5$ kpc ($\sim R_e$) is well constrained by the data, but the mass within $1$ kpc is degenerate with the chosen model.}
  \label{fig:vcirc}
\end{figure}

As demonstrated in Figure~\ref{fig:vcirc}, the dynamical mass profile is best constrained at the maximum radius of the kinematic tracers ($\sim 5$ kpc), with $M_\mathrm{dyn}(< 5 \ \mathrm{kpc}) = 3.4^{+0.5}_{-0.4}(\pm 0.1)\E{9} \ M_\odot$, where the systematic uncertainty (in parentheses) comes from the standard deviation between the four models.
 
Figure~\ref{fig:vcirc} also demonstrates the systematic effect that the choice of halo model has on the inferred circular velocity profile, with both CDM and FDM $\abg$ profiles preferring more massive halos than their associated NFW models by $\sim 0.5$ dex.
This is to be expected, as the cored $\abg$ models put less mass in the inner region (where we have kinematic constraints) compared to NFW models of the same halo mass.
The differences in inferred halo mass between halo models are consistent within the statistical uncertainties from the spread in the posterior distributions, and the deviations indicates the difficulty in robustly extrapolating halo masses out to spatial scales where we lack data.

The analysis of higher order LOS velocity moments (e.g., kurtosis) may help in distinguishing cuspy density profiles (NFW) from shallower cored profiles ($\abg$), as discussed by \citetalias{vanDokkum2019a}.
The high value of $h_4$ measured for the Dragonfly~44 stellar kinematic data favors the $\abg$ model.
However since $h_4$ is more susceptible to systematic biases than the velocity dispersion, we remain largely agnostic about which halo model (and hence which associated value for the halo mass) is correct.

\begin{figure}
  \centering
  \includegraphics[width=\linewidth]{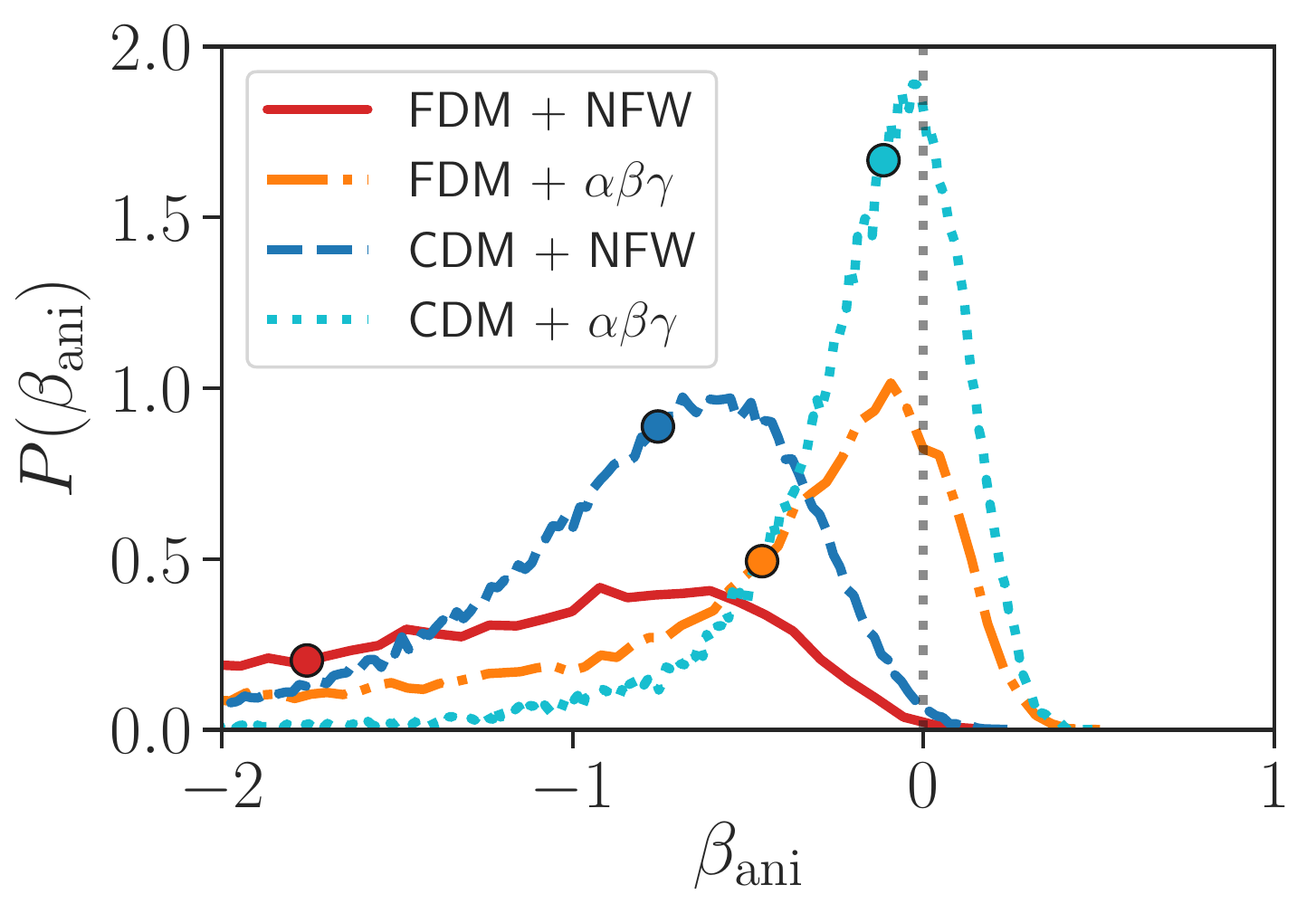}
  \caption{Posterior distributions of the orbital anisotropy parameter for the FDM halo models, compared with their CDM halo counterparts.
    The median of each distribution is marked by circles.
    The NFW models (both for CDM and FDM) prefer tangentially biased orbits ($\beta_\mathrm{ani} < 0$), with the tail of the distributions extending to the prior bound at $\tilde{\beta}_\mathrm{ani} = -1.5$ ($\beta_\mathrm{ani} = -30.6$).
    The $\abg$ models are consistent with isotropic orbits ($\beta_\mathrm{ani} = 0$, shown by the gray dotted line), but all of the posterior distributions are skewed in the direction of tangential anisotropy.}
  \label{fig:anisotropy}
\end{figure}

As expected, the choice of CDM or FDM models has the most impact on the inner mass profile, with FDM models allowing a $\sim 10^9~M_\odot$ core within 1 kpc.
The inner mass distribution is degenerate with both the chosen model and the orbital anisotropy (see Figure~\ref{fig:anisotropy}), with the FDM models preferring more DM inside of 1 kpc and slightly more tangentially-biased orbits.
The primary modeling systematic affecting the anisotropy distribution however is the outer DM profile (NFW or $\abg$), with the NFW model preferring tangential orbits $\beta_\mathrm{ani} \sim -0.8$ and the $\abg$ model preferring isotropic orbits.
We note that models with tangentially-biased orbits will hide the signal of the $v_\mathrm{circ}$ soliton bump when projecting to the LOS velocity dispersion.

\begin{figure}
  \centering
  \includegraphics[width=\linewidth]{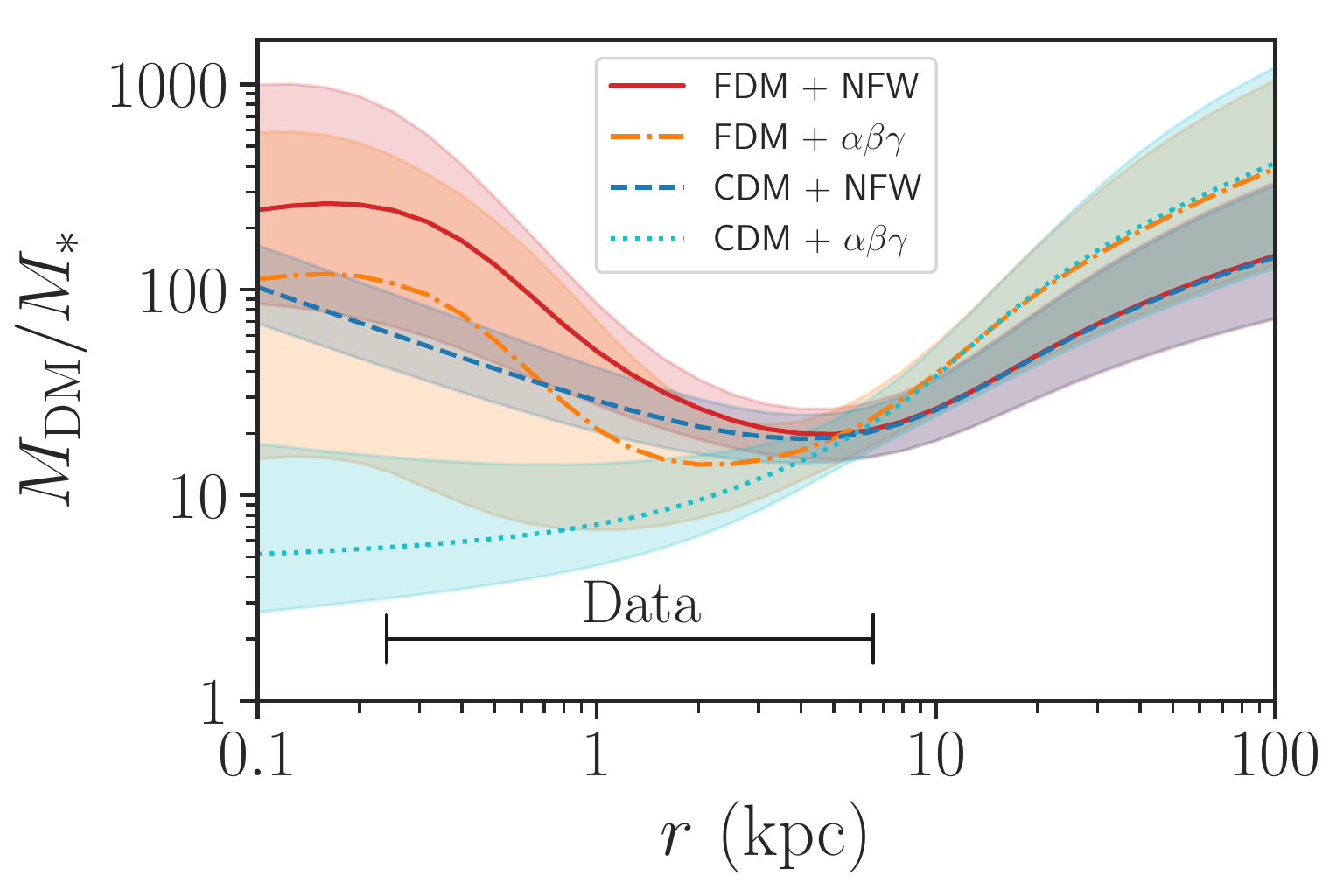}
  \caption{Ratio of DM to stellar mass as a function of radius for the FDM halo models, compared with their CDM halo counterparts.
    The black bar at the bottom indicates the spatial extent of the kinematic data.
    All four models show Dragonfly~44 to be DM-dominated ($M_\mathrm{DM} / M_* > 1$) down to $\sim 0.1$ kpc.}
  \label{fig:dm_ratio}
\end{figure}

Figure~\ref{fig:dm_ratio} shows the ratio of the enclosed (i.e., cumulative) DM mass to stellar mass as a function of radius, and it confirms that Dragonfly~44 is DM-dominated ($M_\mathrm{DM} / M_* > 1$) independently of the considered cosmology (FDM/CDM) or degree of baryonic impacts (NFW/$\abg$), down to the smallest spatial scales probed by the data.
As such, our inference on the mass-to-light ratio, $\Upsilon_*$, is consistent with our chosen prior.
With our chosen prior of $\log \Upsilon_* \sim 0.176 \pm 0.1$, $M_\mathrm{DM} / M_* \sim 20$ at $r = 5$~kpc, independently of the chosen mass model.

\begin{figure}
  \centering
  \includegraphics[width=\linewidth]{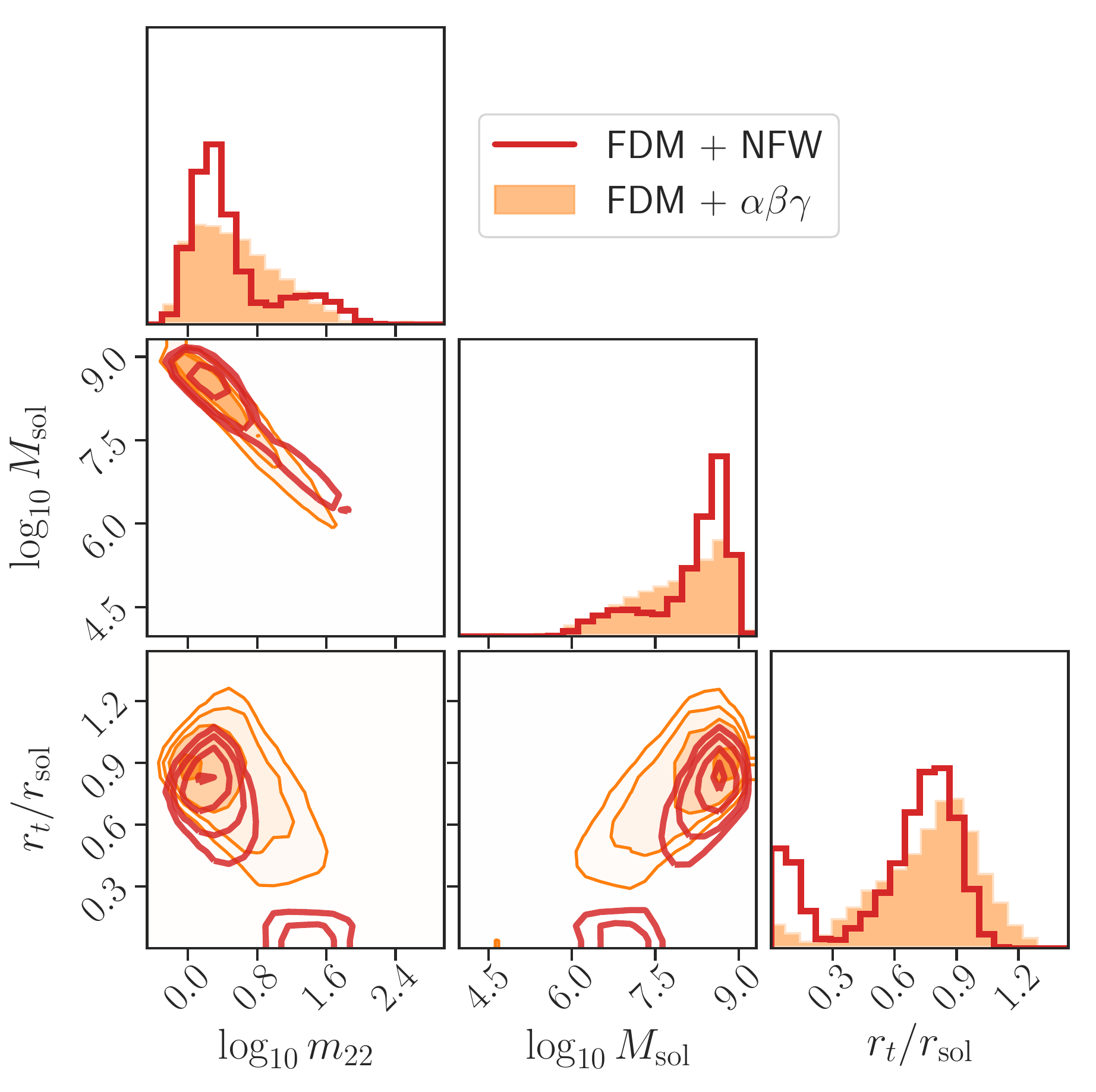}  
  \caption{Posterior distributions for the scalar field mass in $10^{-22} \eV$, the mass within the soliton core (in $M_\odot$), and the ratio of the transition radius to the soliton core radius for the NFW (red, unfilled histograms) and $\abg$ (orange filled histograms) halo models.  
    The FDM constraints are broadly similar between the two halo models, with both models favoring a core of mass $\sim 10^{9}~M_\odot$.
    Both models show a mode in $m_{22}$ of $\sim 2$, with a broad posterior tail towards higher $m_{22}$ values.
    For the NFW model, we see a second mode at high $m_{22}$, corresponding to a negligible core mass and hence a near-zero transition radius.}
  \label{fig:soliton_parameters}
\end{figure}

Looking at just the two FDM models, we see that they are consistent in their posterior soliton parameter distributions.  
Figure~\ref{fig:soliton_parameters} shows the covariance between the scalar field mass, the total mass within the soliton core, and the ratio of the transition radius to the soliton scale radius.  
The modes of the distributions for both NFW and $\abg$ models have a $\sim 10^9 \ M_\odot$ soliton core with a size of $\sim 0.6$ kpc.  
We find a less likely second peak in the posterior distribution for the NFW model, towards a more massive scalar field ($m_{22} \sim 10$).  
This region has a soliton core with mass of $\sim 10^7 \ M_\odot$ that rapidly transitions to the outer NFW halo profile.  
Thus, this second peak corresponds to models for which the DM scalar field is too massive to create a dynamically significant core on spatial scales probed by our data.
For the $\abg$ model, this region of parameter space has a similar posterior density, but this manifests as a long tail towards higher scalar field masses rather than as a discrete second mode.

While the observable velocity dispersion of the FDM models will approach that of the CDM models in the limit as $m_{22} \rightarrow \infty$ (see the bottom right panel of Figure~\ref{fig:cartoon}), we caution that this does \emph{not} mean that the bounded $m_{22}$ posterior distribution favors FDM over CDM.
Rather, as discussed in the beginning of this section, we need to statistically account for the additional model freedom that the introduction of the soliton parameters provide.

\section{Discussion}\label{sec:discussion}

We now focus on the question of whether or not the stellar dynamics of Dragonfly~44 are consistent with the FDM hypothesis and other constraints on FDM.
We find qualitatively similar FDM constraints for both the NFW and $\abg$ models (see Figure~\ref{fig:soliton_parameters}), and so for the sake of simplicity we focus on the FDM + $\abg$ model.

\subsection{Scalar field mass}

We find the DM scalar field mass to be $m_{22} = 3.3^{+10.3}_{-2.1}$.

\begin{figure}
  \centering
  \includegraphics[width=\linewidth]{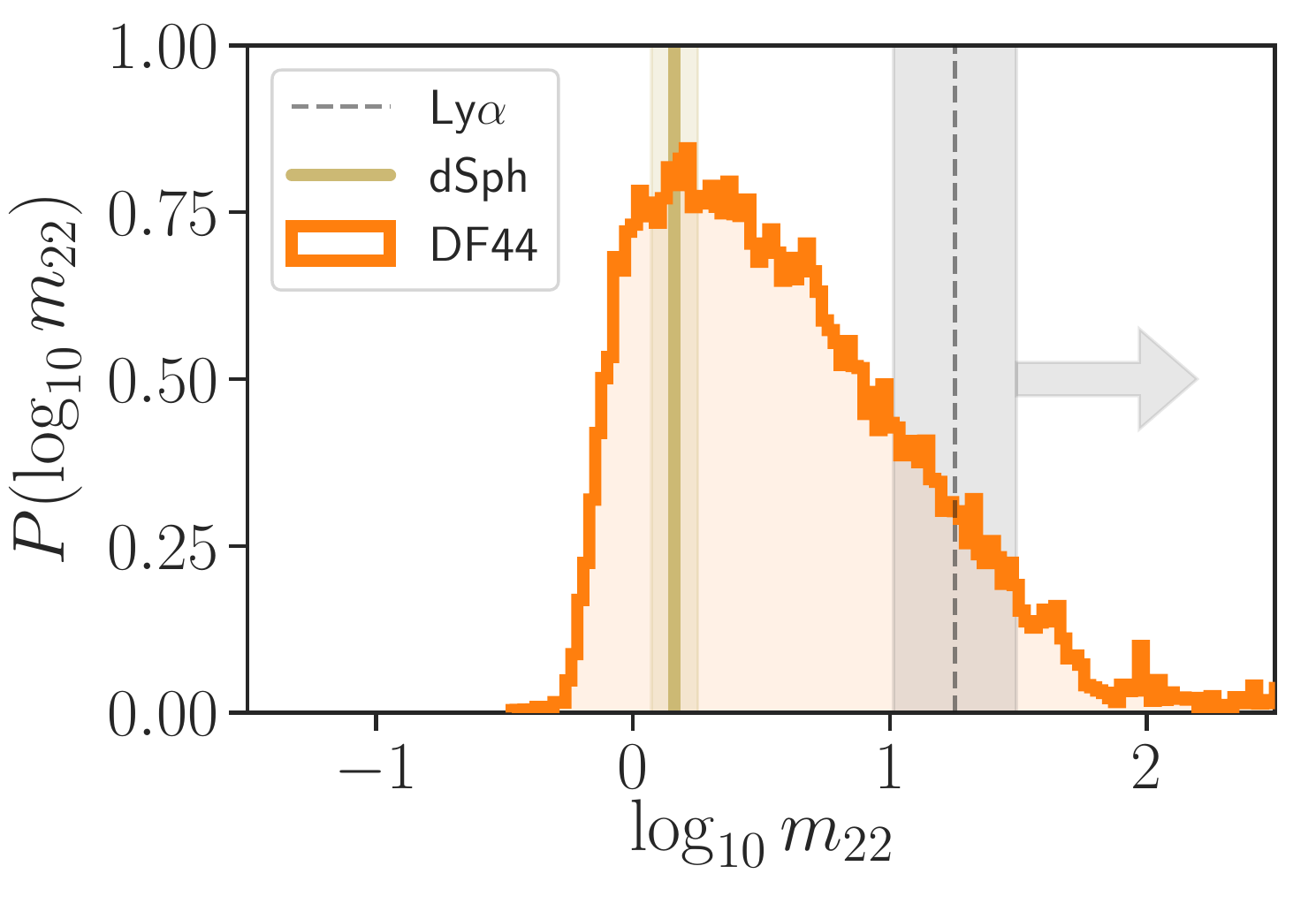}
  \caption{Posterior distributions of $m_{22}$ from Dragonfly~44 (orange histogram) compared with constraints from the literature.
    A lower bound of $m_{22} \gtrsim 20$ from modeling of the Ly$\alpha$ forest (see sources in text) is shown by the gray dashed line, with the gray shaded region showing the range of lower bounds found in the literature.
    The constraint from dSph galaxies \citep{Chen2017} is shown by the yellow solid line.
    We see that both inferences on $m_{22}$ from Dragonfly~44 are consistent with the dSph constraints, but they are in tension with the Ly$\alpha$ constraint.
    Only $\sim 10\%$ of samples lie to the right of the Ly$\alpha$ lower bound.}
  \label{fig:m22}
\end{figure}

Figure~\ref{fig:m22} shows this range in the context of other observational constraints on the scalar field mass.
The values we find for $m_{22}$ are similar to those for the Local Group dSph galaxies from the study of \cite{Chen2017}, who found $m_{22}~\sim~1.8$.
\cite{Gonzalez-Morales2017} found a similar value ($m_{22} \sim 2.4$) from Jeans modeling of the same data, but they cautioned that the orbital anisotropy degeneracy could cause the scalar field mass inference to be biased high.
Instead of using this Jeans analysis, they advocated instead for using mass estimators with multiple stellar subpopulations \citep[e.g.,][]{Walker2011}, for which they derived an upper bound of $m_{22} < 0.4$.

Recent work by multiple authors \citep[e.g.,][]{Irsic2017, Armengaud2017, Kobayashi2017, Nori2019} have used the Ly$\alpha$ forest power spectrum to test FDM.
Less massive FDM particles would result in stronger deviations from $\Lambda$CDM at small spatial scales; thus these studies infer lower bounds on the scalar field mass, with $m_{22}$ values ranging from $7$ to $30$.  

There are a large number of modeling assumptions that go into this lower bound, ranging from the temperature evolution of the intergalactic medium during reionization \citep[e.g.,][]{Garzilli2017} to different priors on cosmological parameters.
In addition, \cite{Desjacques2018} found that even a relatively small self-interaction term in FDM can lead to instabilities that result in notable differences (with respect to CDM) in the cosmic web, complicating the interpretation of Ly$\alpha$ forest clustering.
Thus, it remains uncertain to what degree the FDM constraints from galaxy dynamics and the Ly$\alpha$ power spectrum are in tension with one another.

\subsection{Core size}\label{sec:discussion-core}

The core sizes of soliton halos are predicted to scale with halo mass and scalar field mass as $r_\sol{} \propto m_{22}^{-1} M_h^{-1/3}$.
We can see this by considering the following relations,
\begin{align}
  \label{eq:scalings}
  r_\mathrm{core} &\propto (m v)^{-1} \nonumber \\
  v &\propto \left(\frac{M_h}{r_h}\right)^{1/2}  \\
  r_h &\propto M_h^{1/3} \nonumber
\end{align}
where the first one is from the de Broglie wavelength of the scalar field, the second relation comes from the virial theorem, and the third one comes from the definition of the halo virial radius.  Indeed, inserting relevant constants, we can recover within order unity the scaling relation found from FDM simulations \citep{Schive2014}:

\begin{equation}
  \label{eq:coresize}
  \frac{r_\sol{}}{\mathrm{kpc}} = 5.304 \left(\frac{M_h}{10^{9} M_\odot}\right)^{-1/3} m_{22}^{-1} \ . \end{equation}

\begin{figure}
  \centering
  \includegraphics[width=\linewidth]{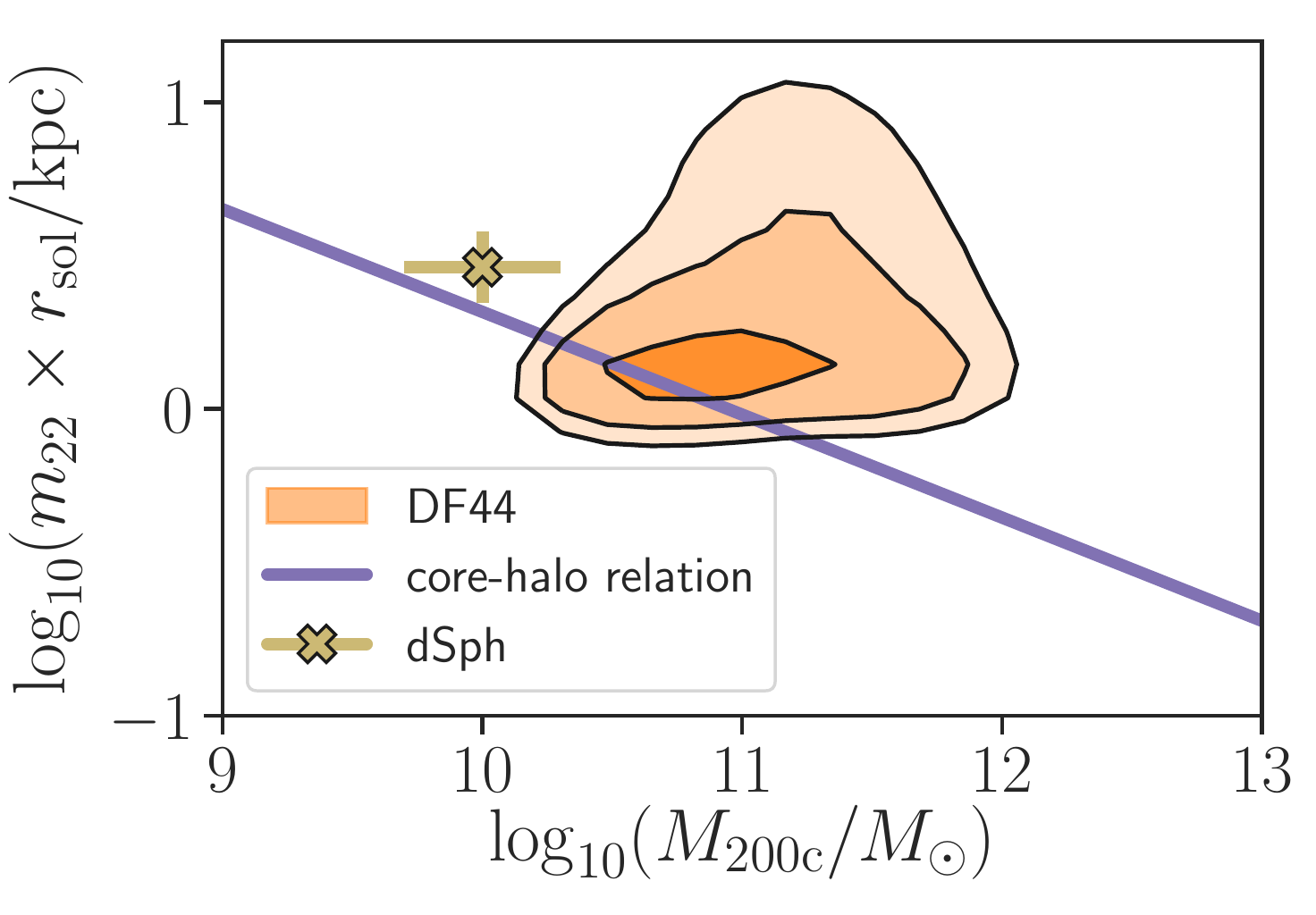}
  \caption{Posterior distribution of $M_\vir$ and $m_{22}\times r_\sol$ for Dragonfly~44 compared to the expected scaling relation.
    The violet line shows the functional relationship between halo mass and core size predicted by \cite{Schive2014}.
    The yellow $\times$ shows the inferred core size from dSph galaxies \citep{Chen2017}.
    There is a broad range of allowed core sizes, but the mode of the distribution is consistent with the expected scaling relation.}
  \label{fig:core}
\end{figure}

We could in principle use Equation~\ref{eq:coresize} as an informative prior on $r_\sol$, which would result in stronger constraints on $m_{22}$.
However, since we let $r_\sol{}$ be a free parameter in our modeling of FDM halos, Equation~\ref{eq:coresize} acts as an additional consistency test for the model.
Figure~\ref{fig:core} shows the posterior distribution of the core size (multiplied by the scalar field mass to remove its associated scaling) and the halo mass.
The mode of the posterior is well-matched to this relation.
In addition, we see that our derived core size for Dragonfly~44 is less than that derived by \cite{Chen2017} for their sample of lower halo mass dSph galaxies, consistent with the direction of the soliton core size--halo mass scaling relation.

\subsection{Transition radius}\label{sec:discussion-transition}

Another consistency check for our FDM models is the location of the transition from the inner soliton profile to the outer CDM-like profile ($r_t$ from Equation~\ref{eq:density}).
For the outer $\abg$ profile, we infer $r_t = 0.5^{+0.4}_{-0.2}$~kpc and $r_t / r_\sol = 0.8^{+0.2}_{-0.3}$.
As shown in Fig.~\ref{fig:soliton_parameters}, these values are similar for the NFW model.

Using simulations of merging FDM halos, \cite{Mocz2017} interpreted this transition radius as the location where the energy density due to quantum pressure is equal to the classical kinetic energy density.  They found this transition radius to occur at $r_t \sim 3.5 \ r_c$ ($\sim 1 \ r_\sol{}$).

Recent work by \cite{Robles2019} identified a plausible range for this ratio of the transition radius to the soliton core radius.
The maximum of this value is set by the requirement that the radius of the peak of the circular velocity profile is less than the virial radius.
The corresponding minimum of this transition ratio is set by either the requirement of a local maximum in the circular velocity profile (for halos $\lesssim 10^{11} \ M_\odot$) or by the need for the peak of the velocity profile in the FDM halo to be less than that of the corresponding CDM halo (for more massive halos).
For a halo of mass $\sim 10^{11} \ M_\odot$, these requirements translate to $0.6 \lesssim r_t / r_\sol{} \lesssim 1.2$.

\begin{figure}
  \centering
  \includegraphics[width=\linewidth]{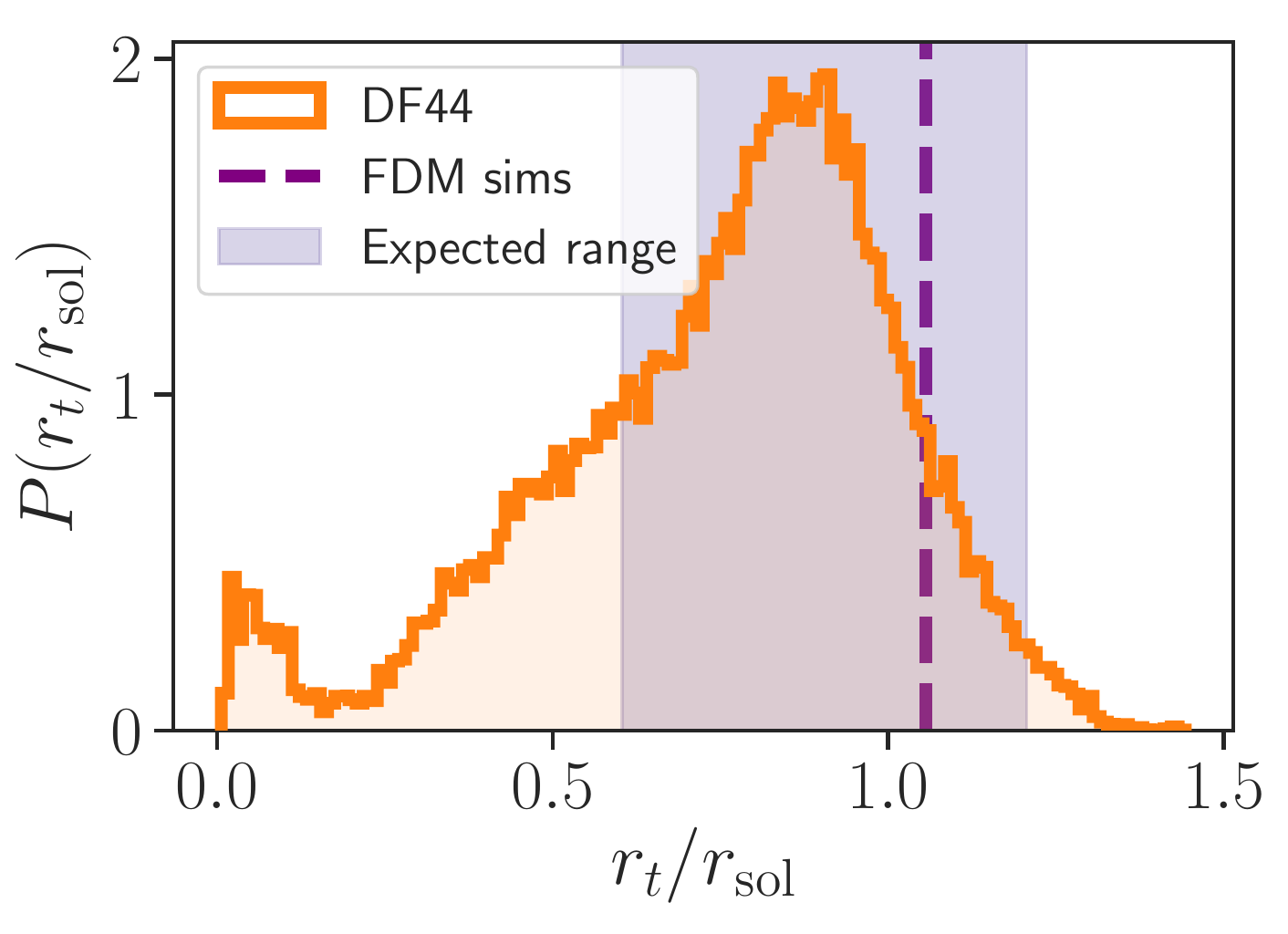}  
  \caption{Posterior distribution of the ratio of the transition radius to the soliton scale radius for Dragonfly~44 (orange histogram), compared with the relevant bounds (violet region) for reasonable FDM halos at the inferred halo mass \citep[see][Sec.~2.2]{Robles2019}.
    The dotted violet line shows the approximate value from the FDM simulations of \cite{Mocz2017}.
    Over two-thirds of the posterior mass for Dragonfly~44 is within these bounds, indicating good agreement with FDM predictions.}
  \label{fig:transition}
\end{figure}

These bounds, as well as the posterior distribution for this transition ratio, $r_t / r_\sol$, are shown in Figure~\ref{fig:transition}.
We recall that our definition of the soliton core radius differs from that used by \cite{Robles2019}, requiring a conversion factor of $3.315$.
In addition, we show the same ratio as found in the simulations of \cite{Mocz2017}.
Most of the posterior mass ($\sim 70\%$) as well as the mode of the distribution is inside of these bounds, indicating that the inferred soliton transition radius is in agreement with the constraints for a reasonable FDM halo.

% \subsection{Orbital anisotropy}

% \begin{itemize}
% \item prospects for constraining $\beta$ with higher order moments
% \item would it help?
% \end{itemize}

% \subsection{Caveats}

% \begin{itemize}
% \item dynamical equilibrium
% \item spherical symmetry
% \item radially-invariant anisotropy
% \item M/L ratio
% \item limited understanding of the effect of baryons in the potential of a FDM halo
% \end{itemize}

% The exact shape of the DM-to-stellar mass ratio profile shown in Figure~\ref{fig:dm_ratio} as well as any comparisons to a stellar-to-halo mass relation is dependent on the priors adopted on $\Upsilon_*$.  

% When we try replacing the Gaussian prior on $\log\Upsilon_*$ with a very wide uniform prior over $(-2.5, 2.5)$ for the CDM NFW model, we find that the halos mass remains largely the same.  We thus find an associated upper bound of $\Upsilon_* \lesssim 11 \ M_\odot / L_\odot$.

\subsection{Future work}

\begin{figure}
  \centering
  \includegraphics[width=\linewidth]{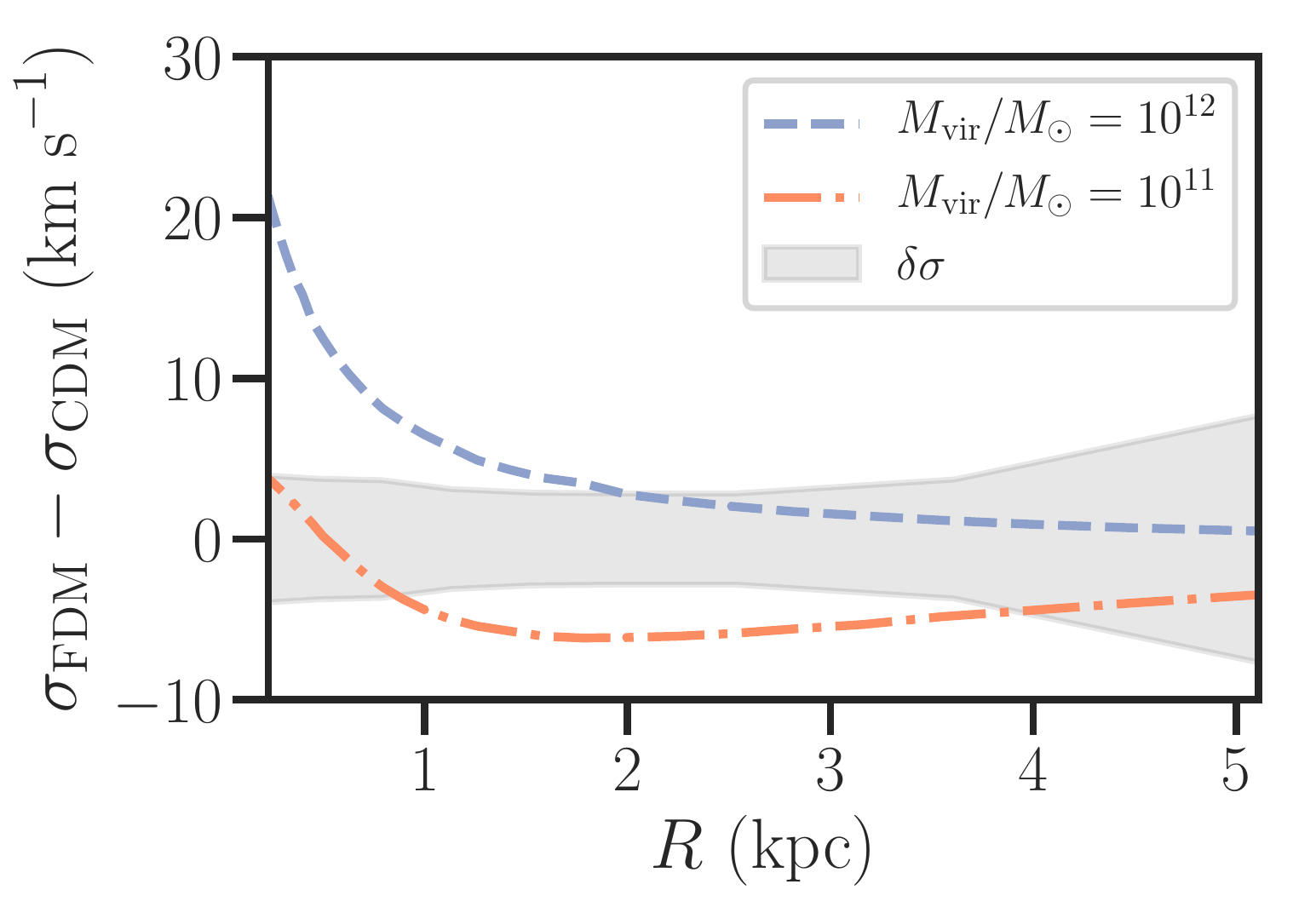}
  \caption{The difference in velocity dispersion between CDM and FDM models, as a function of radius.
    The orange dot-dashed line corresponds to a $10^{11} \ M_\odot$ halo, similar to that inferred for Dragonfly~44.
    The blue dashed line corresponds to a $10^{12} \ M_\odot$ halo, and it demonstrates a much more detectable bump in the velocity dispersion inside of 1 kpc.
    The gray band indicates the observational uncertainties in velocity dispersion for the Dragonfly~44 data.
  Note that this uncertainty region does not take into account the systematic uncertainty in the halo mass profile from the unknown virial mass and concentration.}
  \label{fig:predicted}
\end{figure}

One potentially rewarding area for future work would be testing FDM against galaxies with even higher halo masses than that of Dragonfly~44.
Figure~\ref{fig:predicted} shows that the difference in velocity dispersion between a CDM halo model and a FDM model (both assuming an outer $\abg$ profile) is on the order of the observational uncertainties for a halo mass similar to that of Dragonfly~44.
A $10^{12} \ M_\odot$ FDM halo would be much more readily detected with the current observational error budget.
The field UDG DGSAT~I, with its high velocity dispersion of $\sigma = 56$~\kms \citep{Martin-Navarro2019}, may be one such promising candidate.

As discussed in \citetalias{vanDokkum2019a}, modeling higher order LOSVD moments may help break the mass--anisotropy degeneracy.
Another possibility would be to use the extensive globular star cluster system of some UDGs \citep{vanDokkum2017a} as tracers of the potential.
Such multi-population Jeans modeling can also mitigate the uncertainties from orbital anisotropy \citep[e.g.,][]{Oldham2016, Zhu2016, Wasserman2018}.
% However at the greater distance of Dragonfly~44 ($100$ Mpc vs $\sim~20$~Mpc for previous dynamical studies) the signal-to-noise requirements would demand a much more time-intensive observational campaign.

Most simulation studies of FDM in the literature have not modeled the impact of baryons on the density structure of DM halos (with \citealt{Bar2019} being a notable exception).  
Our crude method for marginalizing over this uncertainty was to try models with the best fit DM profiles from the hydrodynamical simulations of \cite{DiCintio2014a}, which naturally assumed a CDM cosmology.
Stellar feedback may be critical in forming UDGs \citep{DiCintio2017, Chan2018, Jiang2018} and would likely affect the properties of soliton cores in FDM.
Galaxy formation studies with WDM and SIDM \cite[e.g.,][]{DiCintio2017a, Fitts2018, Despali2019} have helped identify better ways of discriminating between available models, and we believe dedicated studies of galaxy formation in a FDM cosmology will be necessary to disentangle the effects of baryonic feedback and new DM physics on the observable DM mass distribution.

\section{Conclusions}\label{sec:conclusions}

We applied equilibrium dynamical models to new spatially-resolved spectroscopy of the integrated starlight of the ultra-diffuse galaxy Dragonfly~44.
We considered FDM halo models in which DM consists of an ultra-light scalar field.

While we were unable to statistically distinguish between our proposed halo mass models, we were able to test the consistency of the FDM halo models.
If we assume a FDM cosmology, the inferred scalar field mass and soliton core size are consistent with a range of FDM predictions, including the core size--halo mass scaling relation and the radius of transition between the soliton core and the outer halo.

The inferred scalar field mass from the Dragonfly~44 data is largely in agreement with other constraints from galaxy dynamics, however it is in tension with results from modeling the Ly$\alpha$ forest power spectrum.
Possible solutions to these disagreements include accounting for any self-interactions in the scalar field or allowing for a mixture of FDM and CDM.
Future work is needed to fully quantify this tension and to determine if FDM is a viable alternative to CDM.

\acknowledgements

We gratefully acknowledge the support of the National Science Foundation via grants
AST-1312376,
AST-1518294,
AST-1613582,
AST-1616598, and 
AST-1616710,
as well as \emph{HST} grant HST-GO-14643.

AJR was supported as a Research Corporation for Science Advancement Cottrell Scholar. 
DAF thanks the ARC for funding via DP160101608.
AV is supported by an NSF Graduate Research Fellowship. 
This work was supported by a NASA Keck PI Data Award, administered by the NASA Exoplanet Science Institute.
Data presented herein were obtained at the W. M. Keck Observatory from telescope time allocated to the National Aeronautics and Space Administration through the agency's scientific partnership with the California Institute of Technology and the University of California.
The Observatory was made possible by the generous financial support of the W. M. Keck Foundation. 
The authors wish to recognize and acknowledge the very significant cultural role and reverence that the summit of Mauna Kea has always had within the indigenous Hawaiian community.
We are most fortunate to have the opportunity to conduct observations from this mountain.

\facilities{Keck:II (KCWI), HST}
\software{numpy \citep{Walt2011}, matplotlib \citep{Hunter2007}, astropy \citep{Astropy2018}, Julia \citep{Bezanson2017}, DifferntialEquations.jl \citep{Rackauckas2017}}

\bibliography{AstroLib}{}
\bibliographystyle{aasjournal}

\appendix

\section{Posterior distributions}\label{sec:appendix-posteriors}

We show the 1D and 2D marginalized posterior distributions for each of the four halo models.
The parameterization and associated units are shown in Table~\ref{tab:parameters}.

\begin{figure}[h]
  \centering
  \includegraphics[width=0.45\linewidth]{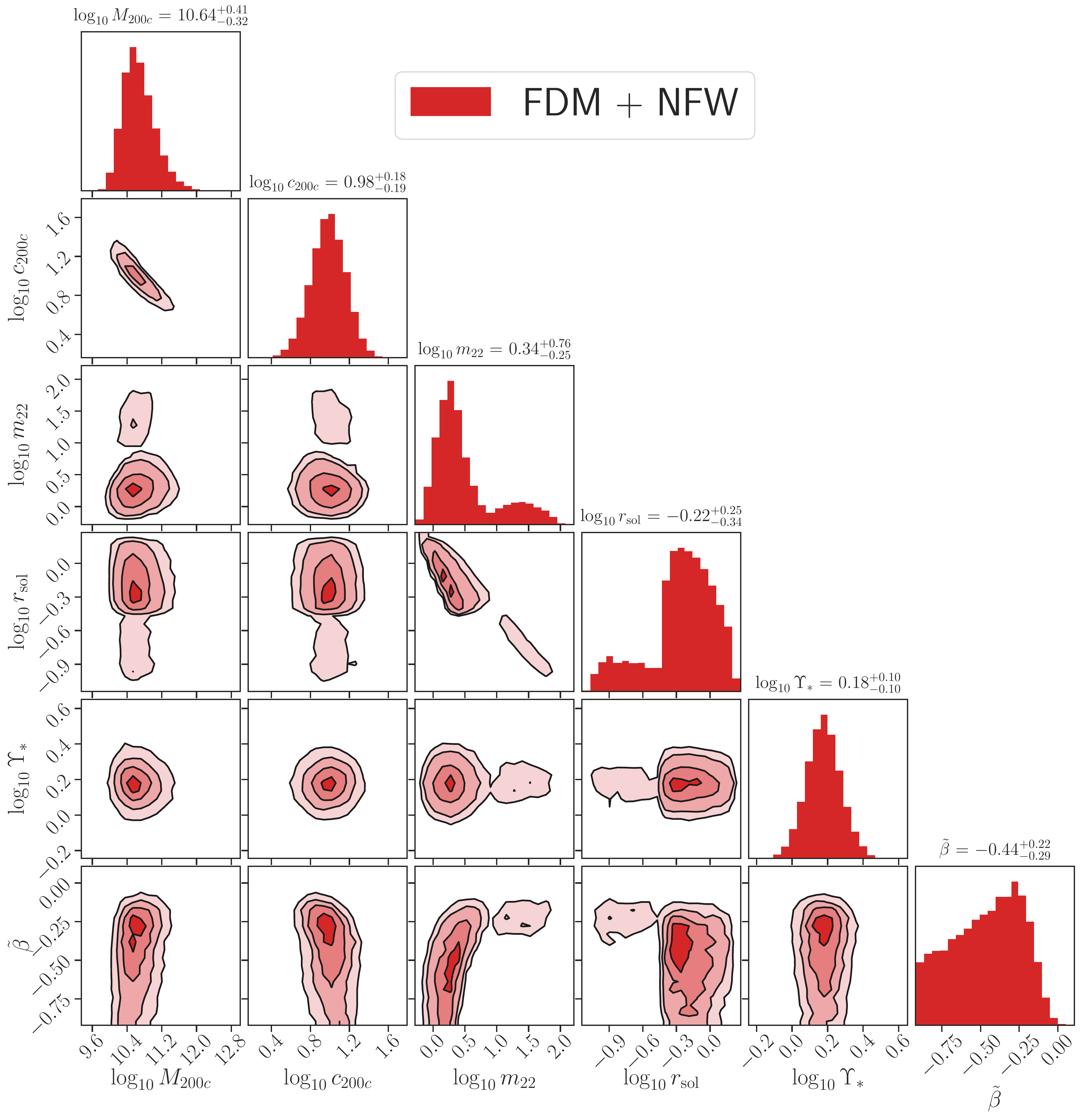}
  \includegraphics[width=0.45\linewidth]{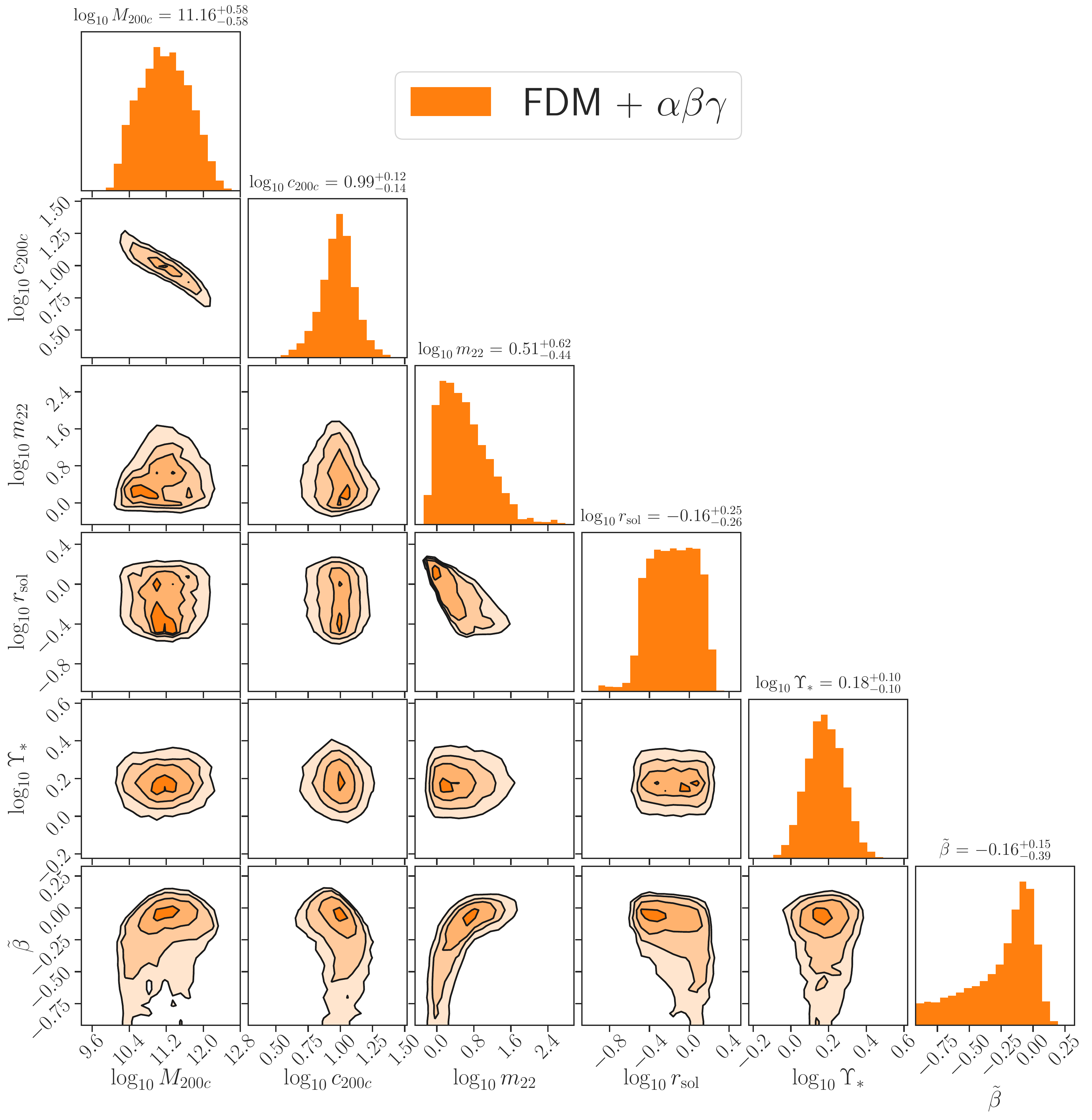}
  \includegraphics[width=0.45\linewidth]{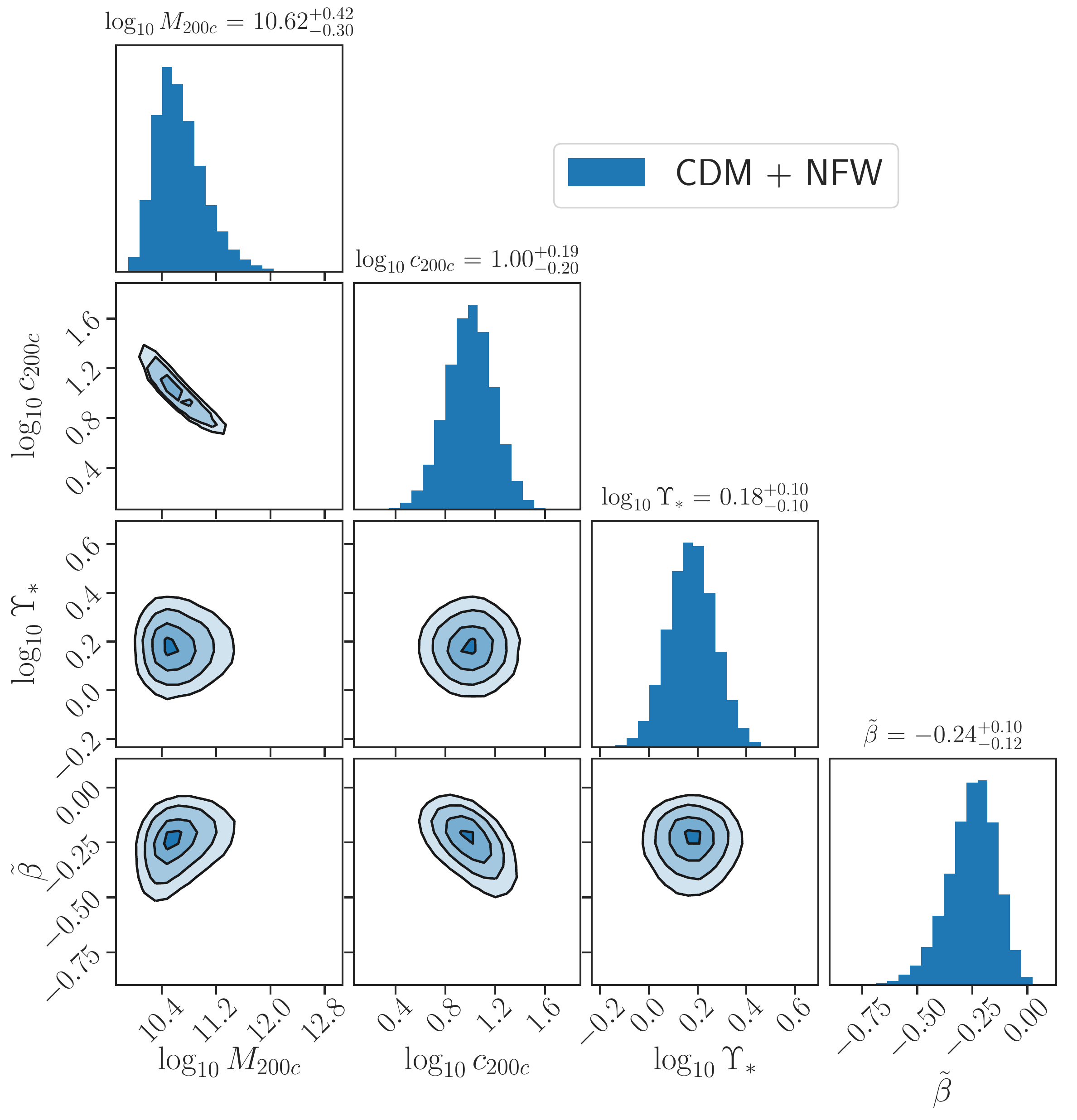}
  \includegraphics[width=0.45\linewidth]{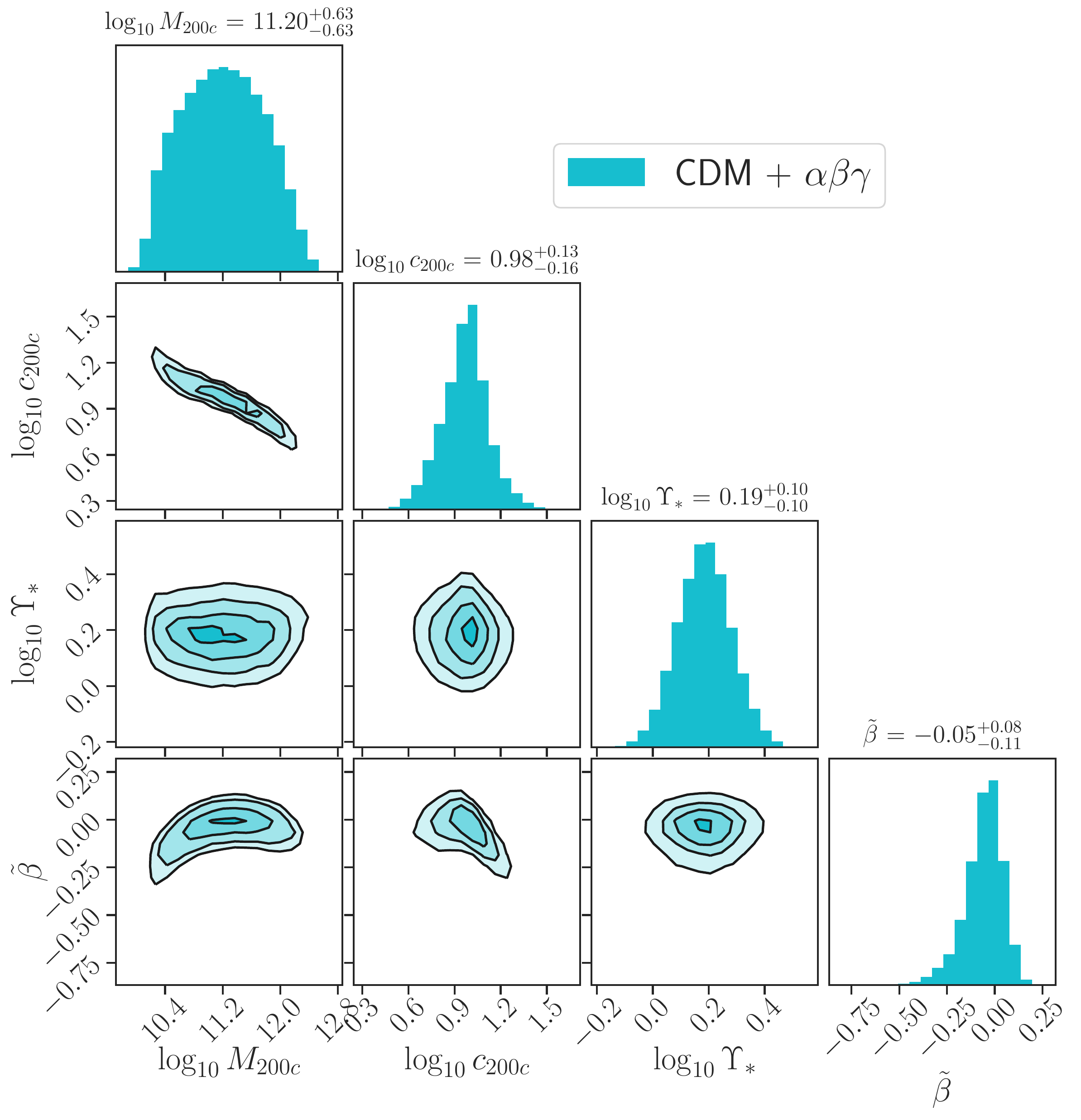}
  \caption{Marginalized posterior distributions for the four halo models. 
    The top panels show the FDM models from this work.
    The bottom panels show the CDM models from \citetalias{vanDokkum2019a}.
    Left panels are for NFW halo profiles, and right panels show the results for the $\abg$ halo profiles.
    Within both top (FDM) panels the parameters are (from left to right, or top to bottom) the log of the halo mass, the log of the halo concentration, the log scalar field mass, the log soliton scale radius, the log of the stellar mass to light ratio, the symmetric parameterization of the anisotropy parameter.
    Contours show iso-density surfaces from 0.5 to 2.0 ``sigma'' levels (for a 2D Gaussian).}
\end{figure}

\end{document}